\documentclass[aps,prx, twocolumn,nofootinbib,superscriptaddress,longbibliography,citeautoscript]{revtex4-1}
\pdfoutput=1

\usepackage[utf8]{inputenc}
\DeclareUnicodeCharacter{00A0}{ }
\usepackage{amsmath}
\usepackage{amssymb}	
\usepackage{graphicx} 
\usepackage{color}
\usepackage[english]{babel}
\usepackage{stmaryrd}
\usepackage{microtype}

\usepackage[percent]{overpic}

\usepackage{hyperref} 
\hypersetup{
	pdftoolbar=true,        		 
	pdfmenubar=true,     		 
	pdffitwindow=false,     	 
	pdfstartview={FitH}, 
	pdftitle={Quantum Quasi-Monte Carlo Technique for Many-Body Perturbative Expansions},  
	pdfauthor={Maček, Dumitrescu, Bertrand, Triggs, Parcollet, Waintal},     		 
	pdfsubject={},   			 
	pdfcreator={},   	        	   	
	pdfproducer={}, 		
	pdfnewwindow=true,      	
	colorlinks=true,       		
	linkcolor=black,          	
	citecolor=blue,        		
	filecolor=magenta,    		
	urlcolor=blue          		
}

\renewcommand{\vec}{\boldsymbol}

\renewcommand{\Re}{\textrm{Re}}
\renewcommand{\Im}{\textrm{Im}}

\newcommand{\ud}{\mathrm{d}}
\newcommand{\vare}{\varepsilon}

\newcommand{\phdag}{{\phantom{\dagger}}}

\newcommand{\header}[1]{\vspace{4pt}\noindent{\textbf{#1.}}}

\newcommand*{\lists}[2]{\left\llbracket \begin{matrix} #1 \\ #2 \end{matrix} \right\rrbracket}

\newcommand{\eqnref}[1]{Eq.~\eqref{#1}}

\begin{document}
\title{Quantum Quasi-Monte Carlo Technique for Many-Body Perturbative Expansions}
\author{Marjan Maček}
\affiliation{Universit\'e Grenoble  Alpes,  CEA,  IRIG-PHELIQS,  38000  Grenoble,  France}
\author{Philipp T.~Dumitrescu}
\email{pdumitrescu@flatironinstitute.org}
\affiliation{Center for Computational Quantum Physics, Flatiron Institute, 162 5th Avenue, New York, NY 10010,  USA}
\author{Corentin Bertrand}
\affiliation{Center for Computational Quantum Physics, Flatiron Institute, 162 5th Avenue, New York, NY 10010,  USA}
\author{Bill Triggs}
\affiliation{Laboratoire Jean Kuntzmann, Université Grenoble Alpes, CNRS, 38401 Grenoble, France}

\author{Olivier Parcollet}
\affiliation{Center for Computational Quantum Physics, Flatiron Institute, 162 5th Avenue, New York, NY 10010,  USA}
\affiliation{Universit\'e Paris-Saclay, CNRS, CEA, Institut de physique th\'eorique, 91191, Gif-sur-Yvette, France}
\author{Xavier Waintal}
\email{xavier.waintal@cea.fr}
\affiliation{Universit\'e Grenoble  Alpes,  CEA,  IRIG-PHELIQS,  38000  Grenoble,  France}
\date{\today} 


\begin{abstract}
High order perturbation theory has seen an unexpected recent revival for
controlled calculations of quantum many-body systems, even at strong coupling.
We adapt integration methods using low-discrepancy sequences to this problem.
They greatly outperform state-of-the-art diagrammatic Monte Carlo. In practical
applications, we show speed-ups of several orders of magnitude with scaling as
fast as $1/N$ in sample number $N$; parametrically faster than $1/\sqrt{N}$ in
Monte Carlo. We illustrate our technique with a solution of the Kondo ridge in
quantum dots, where it allows large parameter sweeps.
\end{abstract}

\maketitle


The exponential complexity of quantum many-body systems is at the heart of many
remarkable phenomena. Advances in correlated materials and recently developed
synthetic quantum systems -- e.g.~atomic gases \cite{Gross_Bloch_2017}, trapped
ions \cite{Blatt_Ross_TrappedIons_2012}, and nanoelectronic devices
\cite{Goldhaber-Gordon1998, Goldhaber-Gordon1998a, Kouwenhoven_Kondo_2000,
Pierre_Science_Kondo_2018} -- have allowed many-body states to be characterized
and controlled with unprecedented precision. The latest of these systems,
quantum computing chips, are highly engineered out-of-equilibrium many-body
systems, where the interacting dynamics performs computational
tasks~\cite{Lukin_51_atoms_2017}. However, our understanding of these many-body
systems is limited by their intrinsic complexity. While uncontrolled
approximations can give insight into possible behaviors, there is a growing
effort to develop controlled, high-precision
methods~\cite{LeblancPRXSimonsColl_2015}, especially ones that apply far from
equilibrium~\cite{Cohen2015, Profumo_1504, Bertrand_1903_series}. These allow us
to make quantitative predictions about the physics of many-body systems and to
uncover qualitatively new effects at strong coupling.

Among theoretical approaches, perturbative expansions in the interaction
strength have seen an unexpected recent revival, in particular using a family of
``diagrammatic'' Quantum Monte Carlo (DiagQMC) methods
\cite{Prokofev_9804,Prokofev_0801,Mishchenko_9910,VanHoucke_1110,Profumo_1504,Wu_1608,Rossi_1612,Chen_1809,Bertrand_1903_series,Bertrand_1903_kernel,Moutenet_1904,Rossi_2001}.
Using various
techniques~\cite{Prokofev_9804,Chen_1809,Bertrand_1903_series,Rossi_2001}, it is
now possible to sum perturbative series beyond their radius of convergence and
thus access strongly correlated regimes. The effects of strong interactions have
been studied in diverse systems, including unitary quantum gases
\cite{VanHoucke_1110}, polarons~\cite{Prokofev_9804}, quantum dots
\cite{Profumo_1504,Bertrand_1903_series,Bertrand_1903_kernel}, and pseudo-gap
metals~\cite{Wu_1608}. 

DiagQMC is currently the preferred strategy for computing series coefficients at
large perturbation order $n$, as this involves integrals of dimension
proportional to $n$ (practically around $5-30$). High dimensional integration is
notoriously difficult, and Monte Carlo provides a robust and flexible solution
with errors that scale as $1/\sqrt{N}$ independently of the dimension; here $N$
is the number of sample points.  

Nonetheless, there has been tremendous progress in integration methods for
problems that lie in-between traditional quadrature (very low dimensions) and
Monte Carlo (high dimensions). In intermediate dimensions (typically 5-200),
`Quasi-Monte Carlo' methods have become well established. These sample the
integrand in a deterministic and structured way that ensures improved uniformity
and better convergence rates. In favorable cases they can achieve error scalings
of $1/N$ or even $1/N^2$, far outperforming traditional Monte Carlo
\cite{Dick_2013,Nuyens_1308,Dick_Pillichshammer_2010,Lecuyer_2018}.

\begin{figure}[b]
    \centering
    \includegraphics[width=\columnwidth]{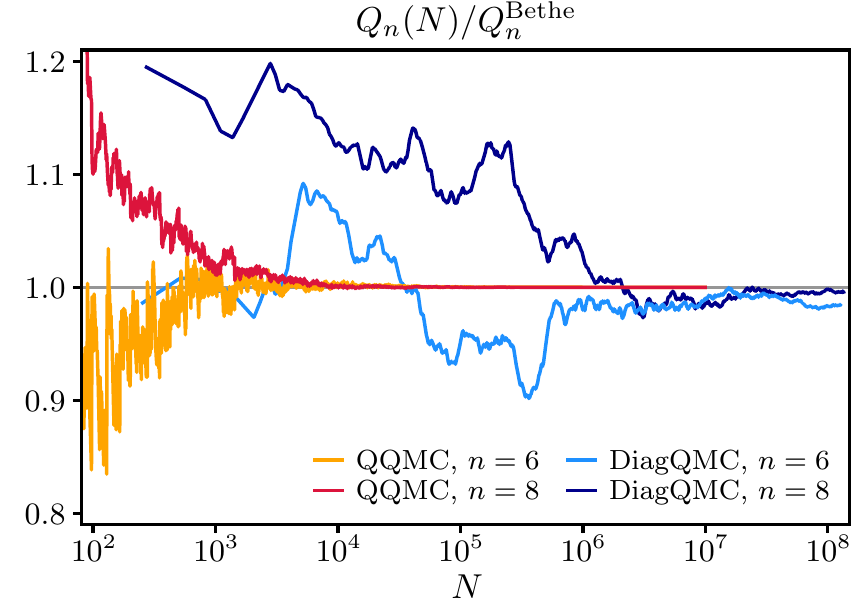}
    \caption{
	\label{fig:Qn-comparison}
	Comparison of the convergence rates for QQMC and DiagQMC. Here $Q_n(N)$ is
	the expansion coefficient of the occupation number of the Anderson impurity
	model at order $n$ as a function of the number of integrand evaluations~$N$.
	Each result is normalized to the exact analytic result
	$Q^{\mathrm{Bethe}}_n$.}
\end{figure}

In this paper we show how to apply these integration techniques to perturbative
expansions for quantum many-body systems.  Our ``Quantum Quasi-Monte Carlo''
(QQMC) approach is broadly applicable. It can be formulated for both equilibrium
and non-equilibrium cases and extended to various lattices and dimensions. Here
we demonstrate it on a quantum dot model and show computational accelerations of
several orders of magnitude compared to state-of-the-art
DiagQMC~\cite{Profumo_1504,Bertrand_1903_series} (Fig.~\ref{fig:Qn-comparison}).
A crucial ingredient of QQMC is the \emph{warping} of the integral. This is a
multi-dimensional change of variables constructed from a model function that
approximates the integrand. We show that a simple model already yields
remarkable results and propose directions for future optimizations. We
demonstrate convergence as fast as $1/N$ in a high-precision benchmark against
an exact Bethe Ansatz solution, to order $n=12$. To illustrate the power of
QQMC, we calculate the finite-bias current through a quantum dot in the Kondo
regime, sweeping electrostatic gating and interaction strength as parameters.
This experimentally relevant calculation was computationally unfeasible for
previous techniques.


\header{Formalism}
In perturbative calculations, an observable $F(U)$ such as a current or
susceptibility is expressed as a power series in the interaction $U$:
\begin{equation} 
F(U) = \sum_{n=0}^\infty F_n U^n,
\end{equation}
where the coefficients $F_n$ are $n$-dimensional integrals
\begin{equation} 
\label{eq:def_int}
F_n =  \int \! \textrm{d}^n\vec{u} \,\, f_n(u_1,u_2,\ldots,u_n).
\end{equation}
The integrands $f_n(\vec{u})$ are time-ordered correlators expressed in terms of
$2^n$ determinants (Wick's theorem), in both Schwinger-Keldysh
\cite{Profumo_1504} and Matsubara formalisms~\cite{Rossi_1612}. The exponential
complexity of evaluating $f_n(\vec{u})$ leads us to seek fast integration
methods. Here the $u_i$ specify the locations of interaction vertices in space
and time.
We present the formalism generally and will specialize to a concrete application
later.

We will perform the integral \eqnref{eq:def_int} by direct sampling using
quasi-Monte Carlo. The crucial step is to warp the integral, i.e.~to make a
change of variables $\vec{u} (\vec{x})$ that maps the hypercube $\vec{x} \in
[0,1]^n$ onto the $\vec{u}$ domain. The integral \eqnref{eq:def_int} becomes
\begin{equation}
\label{eq:warped_int}
F_n = \int_{[0,1]^n} \!\!\ud^n\vec{x}  \,\, f_n\left[\vec{u}(\vec{x})\right]  \left\vert \frac{\partial \vec{u}}{\partial \vec{x}} \right\vert,
\end{equation}
where $\left\vert {\partial \vec{u}}/{\partial \vec{x}} \right\vert$ is the
associated Jacobian. 

The most important property of the warping is to make the function
$\bar{f}_n(\vec{x}) = f_n\left[\vec{u}(\vec{x})\right] \left\vert {\partial
\vec{u}}/{\partial \vec{x}} \right\vert$ as smooth as possible in the new
variables $\vec{x}$. If $f_n$ were positive, the perfect change of variables
would make $\bar{f}_n$ constant and thus trivial to integrate with a single
sample. That would be tantamount to ideal sampling from the distribution
$f_n(\vec{u})$ and it is as challenging as the original integration. Instead, a
judicious warping must provide sufficient smoothing while remaining efficiently
computable.

Mathematically, convergence theorems can only be established for
$\bar{f}_n(\vec{x})$ that belong to specific smooth function spaces, or whose
Fourier coefficients have rapid asymptotic decay properties~\cite{Dick_2013,
Dick_Pillichshammer_2010}. Although we cannot prove that our warped integrands
satisfy assumptions of this kind, in practice we find that the change of
variables are good enough to provide excellent error scaling.

To warp the integral, we consider a positive model function $p_n(\vec{u})$,
which should be viewed as an approximation of $\vert f_n \vert$. The inverse
change of variables $\vec{x}(\vec{u})$ is then defined by (for $1\leq m \leq n$)
\begin{equation}
\label{eq:l_from_u}
x_{m}(u'_{m}, u_{m+1},  \ldots, u_{n})  =  \dfrac{\int_0^{u'_m} \ud u_{m} \int_0^{\infty} \prod_{i=1}^{m-1} \ud u_{i}  \,\, p_n(\vec{u})}
{\int_0^{\infty} \ud u_{m}  \int_0^{\infty}\prod_{i=1}^{m-1} \ud u_{i}   \,\, p_n(\vec{u})}
\end{equation}
Here we adopt a case where $u_i$ is defined on the interval $[0,\infty)$. Since
$x_m(\vec{u})$ only depends on $u_m, \ldots, u_{n}$,  the Jacobian   
is $\left\vert {\partial \vec{u}}/{\partial \vec{x}} \right\vert = [ \int \ud
\vec{u}\,\, p_n(\vec{u})] / p_n(\vec{u})$ (see Appendix~\ref{app:jac}). In
quasi-Monte Carlo, the integral \eqnref{eq:warped_int} is approximated by a sum
over the first $N$ points of a low-discrepancy sequence $\vec{\bar x_i}$. This
is a deterministic sequence of points with specific properties that uniformly
samples the hypercube~\cite{Dick_2013, Dick_Pillichshammer_2010}. We have 
\begin{equation}
\label{eq:warped_int2}
F_n \approx F_n (N) =  \frac{\mathcal{C}}{N} \sum_{i=0}^N \dfrac {f_n\left[\vec{u}(\vec{\bar x_i})\right]}  {p_n\left[\vec{u}(\vec{\bar x_i})\right]}    
\end{equation}
where $\mathcal{C} = \int \ud \vec{u}\,\, p_n(\vec{u})$ is a constant. Here we
use a Sobol' sequence~\cite{Sobol_1967,Kuo_1606} to obtain $\vec{\bar x_i}$.

The model function $p_n(\vec{u})$ should have two key properties. First, it
should approximate $\vert f_n(\vec{u}) \vert$ well. Second, its form should be
simple enough for the partial integrals \eqnref{eq:l_from_u} to be evaluated
exactly and quickly. This allows the reciprocal function $\vec{u}(\vec{x})$ to
be computed by first inverting the one-dimensional function $x_n(u_n)$, then
inverting $x_{n-1}(u_n,u_{n-1})$ for fixed $u_n$, and so on  (see
Appendix~\ref{app:jac}).

Many classes of model functions are possible, as discussed later. This paper
applies the method to impurity models, using a real-time Schwinger-Keldysh
formalism, in which the $u_i$ are the times of the interaction vertices. We
consider the simple form
\begin{equation}
\label{eq:warper1d}
p_n(\vec{u}) = \prod_{i = 1}^{n} h^{(i)}\bigl( u_{i-1} - u_{i} \bigr).
\end{equation}
with $0 < u_n < u_{n-1} < \ldots < u_1 < u_0$. Here $u_0 = t$ is defined to be
the measurement time and the $h^{(i)}$ are positive scalar functions. (They may
depend on $n$,  but we omit this index). The factored structure allows
\eqnref{eq:l_from_u} to be inverted rapidly (see Appendix~\ref{app:jac}).

\header{Anderson Impurity}
We illustrate our method on the Anderson impurity model coupled to two leads.
This is the canonical model for a quantum dot with Coulomb repulsion and the
associated Kondo effect. It has been realized in many nanoelectronic
experiments~\cite{Goldhaber-Gordon1998a,Goldhaber-Gordon1998,
Kouwenhoven_Kondo_2000,Pierre_Science_Kondo_2018}. Importantly, some quantities
including the electron occupation on the dot $Q$ can be computed analytically in
the universal limit with the Bethe ansatz~\cite{Tsvelick_1983_R,Okiji_1984}.
This provides us with a high-precision benchmark for QQMC at any perturbation
order $n$.

We consider an infinite one-dimensional chain with the impurity at site $i=0$.
The non-interacting Hamiltonian is
$H_0 = \sum_{i,\sigma} (\gamma_i c^{\dag}_{i,\sigma} c^{\phdag}_{i+1,\sigma} +
\mathrm{H.c.}) + \vare_d  \sum_\sigma c^{\dag}_{0\sigma} c^{\phdag}_{0\sigma}$,
where $\sigma = \uparrow, \downarrow$ is the electronic spin and $\vare_d$
represents a capacitive gate coupled to the dot. The local Coulomb repulsion is
$H_{\mathrm{int}} = U c^{\dag}_{0\uparrow} c^{\phdag}_{0\uparrow}
c^{\dag}_{0\downarrow} c^{\phdag}_{0\downarrow}$. The electron tunneling between
the leads and dot is $\gamma_0=\gamma_{-1}=\gamma$. All other $\gamma_i = D/2$,
corresponding to hopping within the leads; the lead half-bandwidth $D$ is a
constant. We perform the perturbative expansion in powers of $U$ (see
Appendix~\ref{app:manybody}).

\header{Benchmark}
To validate the QQMC method, we consider the special case solved by the Bethe
Ansatz. For this, we set temperature $T=0$, capacitive gate $\vare_d = 0$, and
half-bandwidth $D \rightarrow +\infty$ such that $\Gamma = 4 \gamma^2 / D = 1$
is the unit of energy. The measurement time $t=30/\Gamma$ is sufficiently long
that the system reaches steady-state. We compute the expansion of the occupation
number  
$Q(U) = \langle c^{\dag}_{0\uparrow} c^{\phdag}_{0\uparrow} +
c^{\dag}_{0\downarrow} c^{\phdag}_{0\downarrow}\rangle$. The system is
particle-hole symmetric for $\vare_d = - U  / 2$ so the non-interacting case is
$Q_0 = 1$. For higher-order $Q_n$, particle-hole symmetry is broken, but the
expansion stays in the symmetric regime $(U + 2\vare_d) \ll
\sqrt{U\Gamma}$~\cite{Tsvelick_1983_R}.

\begin{figure}[t]
\centering
  \begin{overpic}[scale=1]{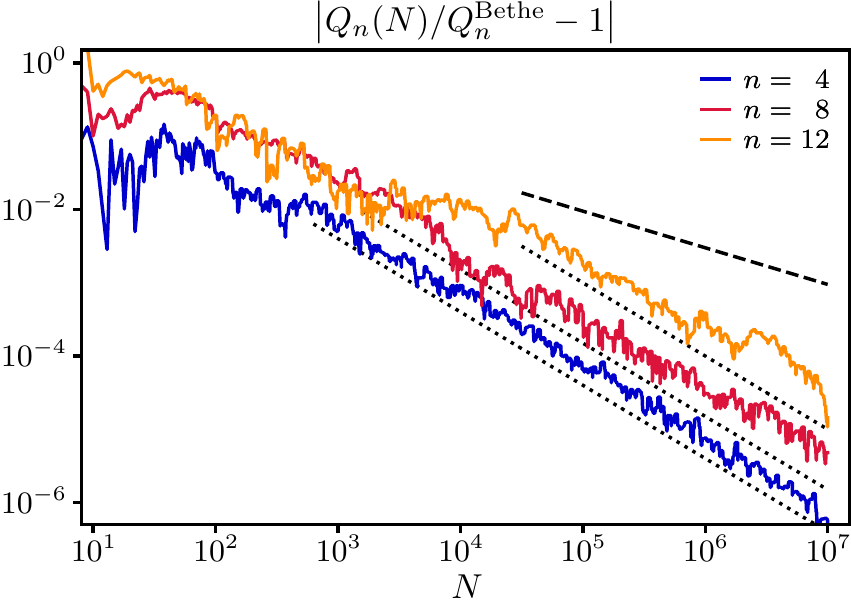}
     \put(10,10){\includegraphics[scale=0.4]{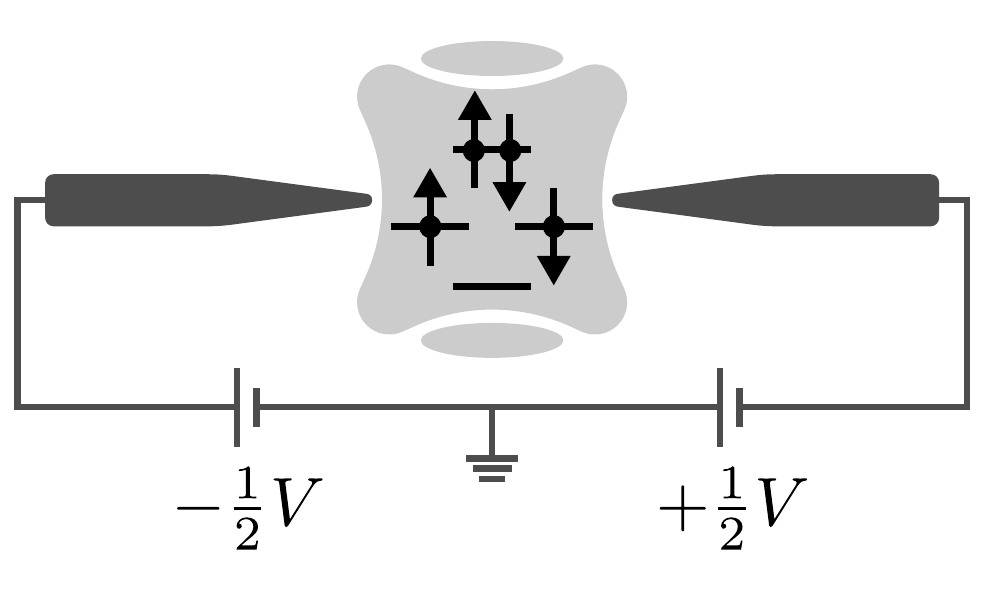}}  
  \end{overpic}
	\caption{\label{fig:Qn-conv-orders} 
	Expansion coefficients $Q_n$ for the Anderson impurity occupation number
	relative to the analytic result $Q^{\mathrm{Bethe}}_n$.  QQMC converges at
	rates close to $1/N$ with the number of integrand evaluations $N$. For
	visibility, the data has been smoothed (see
	Appendix~\ref{app:current_details}). The black lines indicate exact $1/N$
	(dotted) and $1 /\sqrt{N}$ (dashed) convergence. Each run was performed with
	one Sobol' sequence. \emph{Inset:} Cartoon of quantum dot set-up.}
\end{figure}

Figure~\ref{fig:Qn-conv-orders} shows the relative error between $Q_n(N)$ using
QQMC and the exact result $Q^{\mathrm{Bethe}}_n$~(see Appendix~\ref{app:BA}), as
a function of the number of integrand evaluations $N$. 
Following an initial transient, we enter an asymptotic regime in which there is
rapid convergence: for $n=4$ this is consistent with pure $1/N$ while for $n=8,
12$ it is $1/N^\delta$ with $\delta \simeq 0.9, 0.8$. These calculations used
the product model function \eqnref{eq:warper1d} with a single exponential
$h^{(i)}(v_i) = \exp(-v_i/\tau)$, where $\tau = 0.95$. The same set-up was used
in Fig.~\ref{fig:Qn-comparison}. The level of precision that we obtained
revealed limitations in the conventional evaluation of the non-interacting Green
functions, which warranted special consideration~(see
Appendix~\ref{app:manybody}).

It is expected that the convergence rate gradually slows as $n$ increases.
First, the quality of the warping decreases as the disparity between the
increasingly-severe requirements of convergence theory and the behavior of our
integrands grows. This can be mitigated by constructing more expressive model
functions, which we discuss below. Second, for larger $n$ the integrands
generally become more oscillatory. The model functions \eqnref{eq:l_from_u} were
not designed to handle cases with massive cancellation, and this may become a
limiting factor. We will see this effect below for calculations with $\vare_d /
U > 0.5$, although in practice enough orders can be computed accurately to
obtain the desired physical results (see Appendix~\ref{app:manybody}). 

In Quasi-Monte Carlo methods, a standard technique to estimate errors is to
perform computations using \eqnref{eq:warped_int2} with several `randomized'
low-discrepancy sequences
\cite{Dick_2013,Nuyens_1308,Dick_Pillichshammer_2010,Lecuyer_2018} and we use
this method below~(see Appendix~\ref{app:error_calc}).

Having made these technical points, let us reiterate the lessons of
Fig.~\ref{fig:Qn-comparison} and Fig.~\ref{fig:Qn-conv-orders}: (i) QQMC
provides a dramatic speed-up with better asymptotic error scaling than DiagQMC;
(ii) the speed-up persists up to at least order $n=12$, which is what is needed
for practical applications.

\header{Coulomb Diamond}
We now apply QQMC to solve a topical physics problem. We explore the
current-voltage characteristic $I(V)$ across the quantum dot for finite bias and
varying $U$. Since quantum dots are considered promising platforms for building
qubit systems, it is of primary importance to understand how many-body effects
influence their properties, especially the phase coherence.

\begin{figure}[t]
    \centering
    \includegraphics[width=\columnwidth]{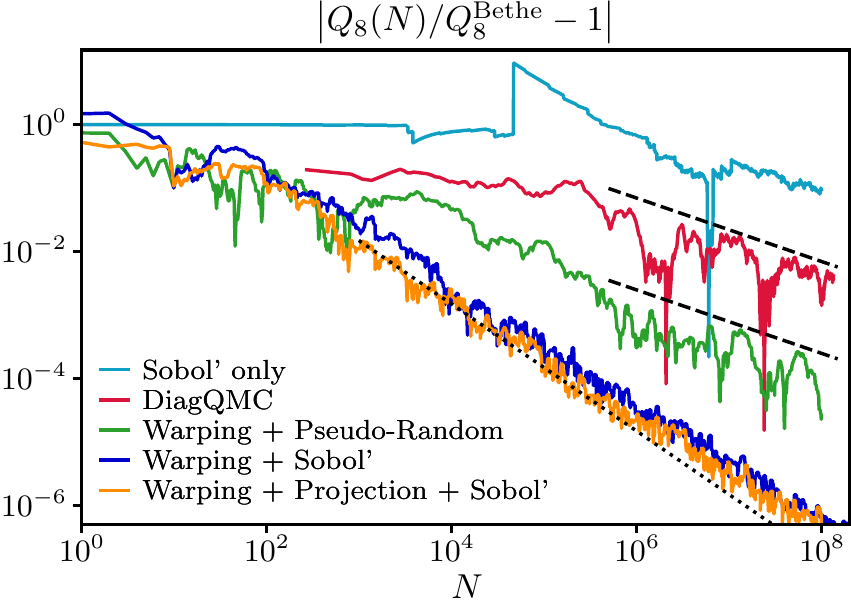}
	\caption{\label{fig:Qn-conv-warpers} Comparison of convergence of $Q_8$ for
	   different methods of integration: evaluating unwarped integrand with a
	   Sobol' sequence (cyan), DiagQMC (red), warped integral sampled with
	   Mersenne Twister pseudo-random numbers (green) or Sobol' sequence (blue).
	   For the warped cases, we used \eqnref{eq:warper1d} with $h^{(i)}(v_i) =
	   \exp(-v_i/\tau), \tau = 0.95$. After an initial warping with exponential
	   functions $\tau = 1.1$, we can apply an additional warping obtained by
	   projection (orange); see Appendix~\ref{app:model_function_details}. For
	   visibility, data (except Sobol' and DiagQMC) has been smoothed in the
	   same way as in Fig.~\ref{fig:Qn-conv-orders}.}
\end{figure}

Quantum dots can be in three different experimentally accessible
regimes~\cite{Tans_1997,Nygard_2000,Liang_2001,Roch_2007}: Fabry-P\'erot (small
$U$), Kondo (intermediate $U$) and Coulomb blockade (large $U$). The
Fabry-P\'erot and Coulomb blockade limits are well described by, respectively,
non-interacting and semi-classical theories; the out-of-equilibrium Kondo regime
is more challenging. Two controlled approaches have recently appeared, but both
are too slow for some applications: the Schwinger-Keldysh DiagQMC used in
Figs.~\ref{fig:Qn-comparison} and \ref{fig:Qn-conv-warpers}~\cite{Profumo_1504,
Bertrand_1903_series} and the real-time inchworm algorithm~\cite{Cohen2015,
Ridley_1907, Krivenko_1904}. QQMC provides the speed and precision to allow
large parameter sweeps, which is mandatory to make good contact with
experiments. In~\cite{Bertrand_1903_series}, some of us studied the Kondo ridge
close to $\vare_d=-U/2$. QQMC allows us to present results scanning the entire
$(U, \vare_d)$ phase diagram, including slowly converging regions with even
numbers of electrons or near the degeneracy points.

\begin{figure}[t]
\centering
\includegraphics[width=\columnwidth]{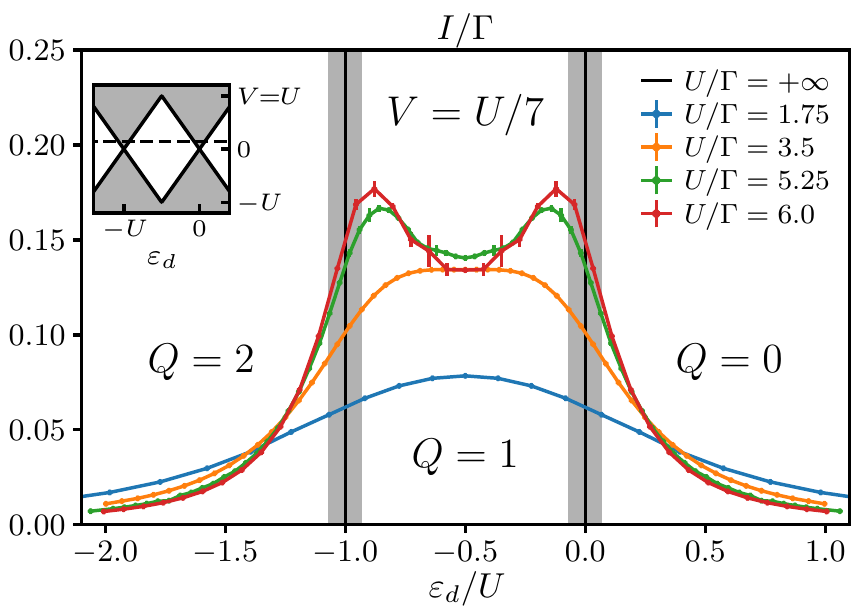}
\caption{\label{fig:current}
Current at finite bias voltage through the Anderson impurity at $T=0$, sweeping
through several interaction regimes. Each point is a different QQMC calculation
up to order $n=10$, including series resummation~\cite{Bertrand_1903_series}.  
The error bars are a combination of integration error and truncation error of
the resummation; the latter dominates. By construction, the data is symmetric
with respect to the particle-hole symmetric point $\vare_d = -U/2$.
\emph{Inset:} Coulomb diamond in the Coulomb blockade picture (large $U$).
Regions where current can flow are shaded grey. The dashed line indicates the
scan shown in the main plot (varying $\vare_d$ for fixed $V/U=1/7$).}
\end{figure}

Figure~\ref{fig:current} (inset) shows a cartoon of the differential conductance
for varying $(\vare_d,V)$ as predicted by Coulomb blockade
theory~\cite{Beenakker_1990} and seen experimentally at low temperatures and
large $U$~\cite{Hofheinz_2007}.  At small bias, the Coulomb blockade forbids
current flows except at two special points: $\vare_d = 0$, where the dot
energies for $Q=0$ and $Q=1$ electrons are degenerate, and $\vare_d = -U$
(likewise for $Q=1,2$). At intermediate $U$, the Kondo effect changes this
picture drastically: the zero-bias Kondo resonance forms in the `forbidden'
region of odd $Q$ and enables current flow.

Figure~\ref{fig:current} shows the current $I$ versus gate voltage $\vare_d$ for
 $V=U/7$ and temperature $T=0$. We choose a finite half-bandwidth $D/\Gamma=20$
 (see Appendix~\ref{app:manybody}). Sweeping the interaction $U/\Gamma$ shows
 several regimes. For $U/\Gamma = 1.75, 3.5$ a current plateau emerges in the
 local moment regime ($Q=1$) due to Kondo resonance formation. The current
 develops new local maxima seen for $U/\Gamma = 5.25, 6.00$. These grow toward
 the Coulomb blockade limit at larger $U$ (black lines); at the same time,
 current around $\vare_d/U = - 0.5$ reaches a maximum and decreases. This is a
 competition between resonance formation and narrowing. At small $U$, the Kondo
 temperature $T_K$  is much larger than the bias $V$ and we are in the linear
 response regime. In this regime near  $\vare_d = -U/2$ we approach perfect
 transmission $I = V/\pi$; see Ref.~\cite{Bertrand_1903_series}. At  larger $U
 \gtrsim 4\Gamma$, $T_K$ decreases exponentially with $U$ and become smaller
 than $V$, leaving the linear response regime. Throughout, as $U$ increases, the
 already-small current in the side regions ($Q=0,2$) is increasingly suppressed.


\header{Model Function}
Let us reexamine the importance of integral warping and model functions.
Figure~\ref{fig:Qn-conv-warpers} shows the convergence of $Q_8(N)$ using
different integration methods; parameters are identical to
Fig.~\ref{fig:Qn-conv-orders}. 
When the integral is evaluated using Sobol' points without warping (`Sobol'
only') the convergence is poor, showing that naively applying low-discrepancy
sequences provides little benefit for these integrands. Next, contrast regular
DiagQMC with the warped integrand using \emph{pseudo}-random numbers. As
expected for pure Monte Carlo approaches, both show $1/\sqrt{N}$ convergence.
Nonetheless, sampling the warped integrand still converges faster than DiagQMC,
despite the fact that the latter uses importance sampling via the Metropolis
algorithm. As anticipated, QQMC using Sobol' points and the model function
\eqnref{eq:warper1d} based on exponential $h^{(i)}$ converges even more rapidly.

How can the model function \eqnref{eq:warper1d} with simple $h^{(i)}$ provide
such dramatic convergence improvements? Our integrands describe physical
correlators that are highly structured and have decaying exponential or
power-law tails; see Appendix~\ref{app:model_function_details} and
Refs.~\cite{Profumo_1504, Bertrand_1903_kernel}. The tail contributions become
ever more important as the dimension increases. The model function properly
describes the long-time asymptotics~(see
Appendix~\ref{app:model_function_details}). We also emphasize the importance of
a well-chosen coordinate system in the model function: the differences of
closest times $v_i = u_{i-1} - u_{i}$ used to parametrize the $h^{(i)}$.

Optimization of the model function should allow further performance gains,
particularly at higher orders $n$. One possibility is to better adapt the
functions $h^{(i)}$ to $f_n$. To illustrate this, we apply a second warping
constructed by sampling points from the first warping. These samples are
projected along the dimensions of $v$ space and smoothed; see
Appendix~\ref{app:model_function_details} for details. As shown in
Fig.~\ref{fig:Qn-conv-warpers}, this optimization reduces the error by a factor
of $\simeq 2$. More importantly, it automatically gives robust convergence
without the need to manually optimize the $\tau$ parameter.

Finally, other families of model functions exist beyond \eqnref{eq:warper1d},
that provide versatile and expressive approximations while still allowing for
fast inversion of \eqnref{eq:l_from_u}. One such family is Matrix Product States
(MPS) or functional tensor-trains~\cite{Schollwoeck_1008, Glasser_1907}, of
which \eqnref{eq:warper1d} is just the simplest case:
\begin{equation}
\label{eq:warperMPS}
p_n(\vec{u}) = h^{(1)}_{a}(v_1) h^{(2)}_{ab}(v_{2}) \cdots h^{(n-1)}_{cd}(v_{n-1}) h^{(n)}_{d}(v_n).
\end{equation}
Here $h^{(i)}_{ab}$ are matrices and repeated indices are summed. Another
promising family is  
$
p_n(\vec{u}) = \prod_{i=1}^{n-1}  \bar h^{(i)}(v_{i+1},v_{i}).
$

\header{Conclusion} 
We have shown how to use sampling techniques based on low-discrepancy sequences
to compute high orders of many-body perturbation theory. Although we cannot show
that the integrands obey the assumptions of formal Quasi-Monte Carlo convergence
theory, practical scaling as fast as $1/N$ is still achievable. This success was
possible due to the warping of the integral based on a model function. Using
benchmarks on exactly solvable quantities in the Anderson impurity model, we
unambiguously validated the convergence of this `Quantum Quasi-Monte Carlo'
(QQMC) method at high-precision. This calculation was about $\sim 10^4$ times
faster than the DiagQMC equivalent.

We can apply the techniques established here to models with interesting strongly
correlated physics in all dimensions, for equilibrium and especially
non-equilibrium situations. For continuum models, the integrands are smooth and
warping should be particularly simple. For lattice models, the discrete
summation may degrade convergence, although this may be addressed with
sufficiently good model functions. QQMC can also be applied to other
diagrammatic expansions, e.g.~in hybridization~\cite{Cohen2015}. Constructing
more expressive model functions should further increase speed and accuracy and
is an ideal application for recent machine learning techniques in quantum
systems.

Finally, we have shown in our calculations that the simple model function
\eqnref{eq:warper1d} captures the behavior of perturbation theory integrands in
asymptotic large-coordinate regions. This is not accidental, but reveals a
simplifying structure of the correlation functions arising from Wick's theorem
that was not previously appreciated in diagrammatic numerical simulations. For
the real-time Schwinger-Keldysh calculations, the contour index means that the
MPS structure \eqnref{eq:warperMPS} is the natural approximation for generic
many-body systems. It can be used as a starting point to efficiently compute and
integrate these functions, even beyond the Monte Carlo or QQMC sampling
discussed here.


\begin{acknowledgments}
We thank N.~Andrei, L.~Greengard, E.M.~Stoudenmire, N.~Wentzell and especially
A.H.~Barnett for helpful discussions. The algorithms in this paper were
implemented using code based on the TRIQS library~\cite{TRIQS2015} and the
QMC-generators library~\cite{Kuo_1606}. The Flatiron Institute is a division of
the Simons Foundation. XW and MM acknowledge funding from the French-Japanese
ANR QCONTROL, E.U.~FET UltraFastNano and FLAG-ERA Gransport.
\end{acknowledgments}

\appendix

\section{Low-Discrepancy Sequences}
\label{app:quasi}

\begin{figure}[b] \centering
    \includegraphics[width=\columnwidth]{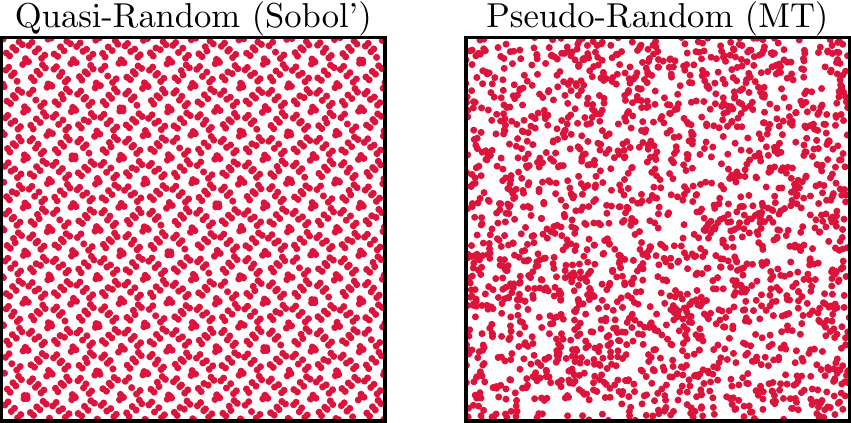}
    \caption{\label{fig:mt_sobol} Sampling the two-dimensional square $[0,1]^2$ with
    quasi-random numbers from a Sobol' sequence (left) and pseudo-random numbers
    from a Mersenne-Twister sequence (right). The Sobol' sequence gives a far more
    uniform sampling.}
\end{figure}

The integration technique for perturbation theory integrals we presented in the
main manuscript is based on so-called `quasi-Monte Carlo' methods, which use
sampling with low-discrepancy sequences. Despite the naming, such quasi-random
sequences are highly structured and deterministic, unlike the pseudo-random
number sequences used in conventional Monte Carlo sampling (see
Fig.~\ref{fig:mt_sobol}). Here we briefly summarize the history of this field of
mathematics as well as key features of these sequences.

The roots of low-discrepancy and quasi-random sequences lie in the ancient field
of Diophantine Approximation -- the theory of approximating sets of real numbers
by rational ones, especially in its modern form pioneered by Roth
\cite{Roth_1954} -- and in Weyl's early efforts to characterize uniformity of
distributions \cite{Weyl_1916}. These methods were applied to practical
multidimensional integration in the late 1950's and 1960's, with contributions
from, among many others, Korobov \cite{Korobov_1957} and Sobol'
\cite{Sobol_1967} in the USSR and Hammersley \cite{Hammersley_1960} (building on
work by van der Corput \cite{Corput_1935}), Halton \cite{Halton_1960} and
Haselgrove \cite{Haselgrove_1961} in the UK. There was even an unsuccessful
attempt to use them as a component of Markov Chain Monte Carlo simulation for
multi-point physical integrals in 1951 \cite{Richtmyer_1952}. Niederreiter
\cite{Niederreiter_1978} surveys much of the early work in this area. Since that
time, quasi-Monte Carlo methods have become standard in many fields including,
for example, computational chemistry \cite{Conroy_1967}. Indeed, convergence
rate improvements analogous to, but smaller than, ours are seen in molecular
excitation computations \cite{Berblinger_1991}. Most of the early applications
had relatively low dimensions $\lesssim 10$. However, the field was rejuvenated
in the late 1990's and early 2000's by the discovery that quasi-Monte Carlo
works unexpectedly well for certain high-dimensional financial integrals
stemming from discretized stochastic partial differential equations
\cite{Paskov_1995}. This was followed by the application of Reproducing Kernel
Hilbert Space theory to explain this success (see e.g.~Ref.~\cite{Wang_2005}),
and the advent of `fast component-by-component' construction techniques to
create optimized quasi-random generators for these problems \cite{Nuyens_2006}.
A survey of this more recent work appears in Ref.~\cite{Dick_2013}.

The main strength of quasi-Monte Carlo relative to Markov Chain Monte Carlo is
its greatly accelerated convergence rate in applications with well-behaved
integrands. Its main disadvantage is its reduced flexibility. Quasi-Monte Carlo
methods require highly symmetric integration domains -- usually hypercubes --
and careful preparation of the integrand. In our approach we implement this
through a change of coordinates and warping. However, perhaps most importantly,
they are intrinsically non-adaptive during a calculation: changing the
distribution during the main sampling run destroys its carefully constructed
uniformity properties, which typically reduces the convergence rate to Monte
Carlo scaling $O(1/\sqrt{N})$. To handle generic integrands whose properties are
unknown in advance, separate runs and analysis are needed before the main
sampling run.

Finally, we note that the theory of quasi-Monte Carlo typically gives worst-case
error bounds, but not average-case ones. This is because no stochastic averaging
is done. An exception is when we perform averaging of random shifts. These
worst-case bound are theoretically tight but often rather pessimistic compared
to observed behavior. For pure Monte Carlo methods, the opposite situation is
true and we typically have average-case but not worst-case bounds.

\section{Summary of the QQMC algorithm}
\label{app:summary}

Below, we summarize the QQMC algorithm. More details on the building blocks are
given in other sections of this supplementary material. Note that unlike usual
diagrammatic Monte Carlo techniques, the coefficients $F_n$ for different orders
$n$ are calculated separately.

\header{Preprocessing steps}

\begin{itemize} 
\item Calculate non-interacting Green functions in real time. These functions
form the basic elements from which the integrands are calculated. They can be
obtained through Fourier transform of the real frequency Green functions which
can be obtained analytically or numerically. 
\item Calculate Warping. Once a model function $p_n(\vec u)$ has been selected
and possibly adapted to the integrand (see projection method below), construct
the mapping $\vec x (\vec u)$ by computing the partial cumulative functions of
the model function $p_n(\vec u)$. 
\item For each component of the mapping $\vec x(\vec u)$, invert it to obtain
the inverse mapping $\vec u (\vec x)$. This can be done, for example, be
interpolating the original mapping $\vec x(\vec u)$ on a sufficiently fine
linear mesh and inverting.
\item Initialize $r=0$.
 \end{itemize}

\header{Main computing loop} 
For each sample point $i\in 1\dots N$,

\begin{itemize} 
\item Generate the low-discrepancy quasi-random vector $\vec x_i$. These vectors
span uniformly the hypercube $[0,1]^n$. 
\item Calculate the corresponding point $\vec u_i = \vec u(\vec x_i)$ in the
original integration space. 
\item Calculate the corresponding value of the integrand $f^{(i)} = f_n(\vec
u_i)$ and the value of the model function $p^{(i)} = p_n(\vec u_i)$. 
\item Update the result $r\rightarrow r + f^{(i)}/p^{(i)}$
 \end{itemize}

\header{Returns} The final estimate of the observable is $F_n \approx r/N$.

\section{Benchmarking at higher orders} 
\label{app:highorders}

\begin{figure}[b] 
    \centering
    \includegraphics[width=\columnwidth]{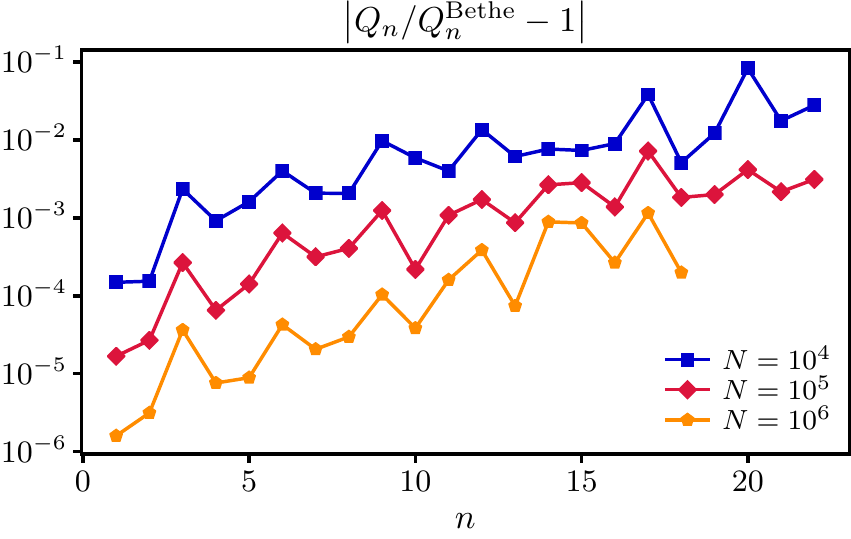}
    \includegraphics[width=\columnwidth]{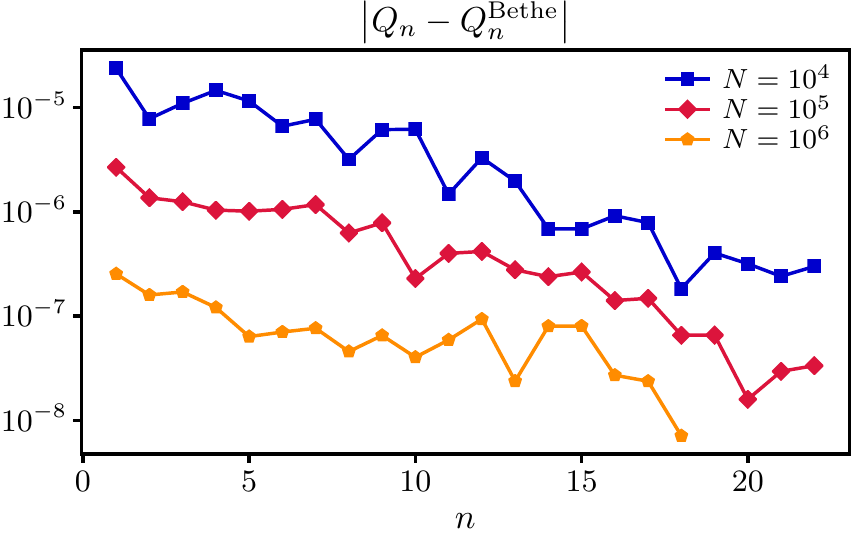} 
    \caption{
    \label{fig:high-order} 
    Relative error (upper) an absolute error (lower) of $Q_n$ versus perturbation
    theory order $n$ for our benchmark calculation against Bethe ansatz results, up
    to $n=22$. Different curves correspond to different numbers of calculated points
    $N$. We used \eqnref{eq:warper1d} with $h^{(i)}(v_i) = \exp(-v_i/\tau)$ with
    $\tau = 0.95$ for $n \leq 15$ and $\tau = 0.90$ for $n \geq 16$. The parameters
    are the same as in Fig.~\ref{fig:Qn-comparison}.}
    \end{figure}

The main manuscript focuses on data for perturbation theory orders $n \lesssim
12$, where QQMC enables calculations with an unprecedented precision of up to
six or seven digits. In this section, we show additional data for a smaller
number of samples $N$. This enables us to calculate much larger orders up to
$n=22$, well beyond previous diagrammatic quantum Monte Carlo techniques. We
recall that the a single evaluation of the integrand $f_n(\vec{u})$ has an
exponential complexity $\sim 2^n$, which has limited previous calculations to
order $n \le 15$ with less than two digits precisions. While $N=10^5$ is not yet
deep into the asymptotic regime, we find that we could reach two to three digits
of accuracy.

The data are presented in Fig.~\ref{fig:high-order} respectively for the
relative error (top panel) and absolute error (lower panel). Although the error
deteriorates with the order $n$, the speed up provided by QQMC more than
compensates for the imperfection of the model function. We find that the
absolute error actually decreases with $n$ as the $Q_n$ gets smaller at large
$n$.

\section{Model Function Properties}
\label{app:jac}

Here we expand on the properties of the change of variables $\vec{x}(\vec{u})$
arising from the model $p_n(\vec{u})$. These were defined in the main text as:
\begin{equation}
\tag{\ref{eq:l_from_u}$'$}\label{eq:l_from_u_app}
x_{m}(u'_{m}, \ldots, u_{n})  =  \dfrac{\int_0^{u'_m}  \ud u_{m} \int_0^{\infty} \prod_{i=1}^{m-1} \ud u_{i}  \,\, p_n(\vec{u})}
{\int_0^{\infty} \ud u_{m}  \int_0^{\infty} \prod_{i=1}^{m-1} \ud u_{i}   \,\, p_n(\vec{u})}
\end{equation}
To understand the structure of \eqnref{eq:l_from_u_app} and how it can be useful
when sampling from $p_n(\vec u)$, let us consider the explicit transformation
for small orders $n$. 

For $n=1$, \eqnref{eq:l_from_u_app} is simply the normalized cumulative function
$x_1(u_1) = \int_0^{u_1} \ud\bar{u}_1 p_1(\bar{u}_1)/\int_0^{\infty} \ud
\bar{u}_1 p_1(\bar{u}_1)$ which upon differentiation gives,
\begin{equation}
dx_1 = \frac{p_1(u_1) \ud u_1}{\int_0^{\infty} \ud  \bar{u}_1 \,\, p_1( \bar{u}_1)}.
\end{equation}
This means that uniformly sampling $x_1$ leads to the sampling of $p(u_1)$. In
practice, one need to invert the cumulative distribution $x_1(u_1)$ which can be
done through interpolation techniques. 

Next, let us consider the procedure for $n=3$ using a model function $p_3(u_1,
u_2, u_3)$. The reverse coordinate transform $\vec{x}(\vec{u})$ is
\begin{align}
\nonumber x_{1}(u_1, u_2, u_3) & =  \frac{\int_0^{u_1}  \ud \bar{u}_{1}\,\, p_3(\bar{u}_1, u_2, u_3)}{
\int_0^{\infty}  \ud \bar{u}_{1}\,\, p_3(\bar{u}_1, u_2, u_3)} \\
x_{2}(u_2, u_3) & =  \frac{\int_0^{u_2}  \ud \bar{u}_{2}\,\, \int_0^{\infty}  \ud \bar{u}_{1}\,\, p_3(\bar{u}_1, \bar{u}_2, u_3)}{
\int_0^{\infty}   \ud \bar{u}_{2}\,\,\int_0^{\infty}   \ud \bar{u}_{1}\,\, p_3(\bar{u}_1, \bar{u}_2, u_3)} \label{eq:app:LFU3}  \\
\nonumber x_{3}(u_3) & =  \frac{\int_0^{u_3}  \ud \bar{u}_{3}  \int_0^{\infty}  \ud \bar{u}_{2} \int_0^{\infty}  \ud \bar{u}_{1}\,\, p_3(\bar{u}_1, \bar{u}_2, \bar{u}_3)}{
\int_0^{\infty}   \ud \bar{u}_{3} \int_0^{\infty}   \ud \bar{u}_{2}\int_0^{\infty}   \ud \bar{u}_{1}\,\, p_3(\bar{u}_1, \bar{u}_2, \bar{u}_3)}
\end{align}
The consecutive coordinate integration, gives the coordinate transformation a
special structure: $x_{3}(u_3)$ does not depend on $u_1, u_2$ and $x_{2}(u_3,
u_2)$ does not depend on $u_1$. This means that the Jacobian matrix for the
reverse coordinate transformation $\vec{x}(\vec{u})$ has an upper-triangular
form
\begin{equation}
\left[ \frac{\partial \vec{x}}{\partial \vec{u}} \right] 
= \left[
\begin{array}{ccc}
({\partial x_1}/{\partial u_1})  & ({\partial x_1}/{\partial u_2})  & ({\partial x_1}/{\partial u_3})   \\
0 & ({\partial x_2}/{\partial u_2})  & ({\partial x_2}/{\partial u_3})  \\
0 & 0 & ({\partial x_3}/{\partial u_3})
\end{array}
\right]
\end{equation}
so that the Jacobian determinant is simply
\begin{equation}
\left\vert \frac{\partial \vec{x}}{\partial \vec{u}} \right\vert = 
\frac{\partial x_1}{\partial u_1} \cdot  \frac{\partial x_2}{\partial u_2}  \cdot \frac{\partial x_3}{\partial u_3}.
\end{equation}
Differentiating \eqnref{eq:app:LFU3}, cancelling common factors and using
$\left\vert {\partial \vec{u}}/{\partial \vec{x}}\right\vert = 1 / \left\vert
{\partial \vec{x}}/{\partial \vec{u}}\right\vert$ gives
\begin{align}
\left\vert \frac{\partial \vec{u}}{\partial \vec{x}} \right\vert
 = &  \frac{
\int_0^{\infty}   \ud u_{3} \int_0^{\infty}   \ud u_{2}\int_0^{\infty}   \ud u_{1}\,\, p_3(u_1, u_2, u_3)}{ p_3(u_1, u_2, u_3)}.
\end{align}

The same procedure straightforwardly generalizes to an arbitrary number of
dimensions $n$. This reproduces the result quoted in main text:
\begin{equation} \label{eq:jac-norm_app}
\left\vert \frac{\partial \vec{u}}{\partial \vec{x}} \right\vert = 
 \frac
{ \int_0^\infty \prod_{i=1}^{n} \ud u_{i}   \,\, p_n(\vec{u})}{ p_n(\vec{u})}.
\end{equation}

In practice, we uniformly sample the hypercube $[0,1]^3$ using Sobol’ sequence
to obtain $(x_1,x_2,x_3)$. From $x_3$ one obtains $u_3$ by inverting the one
dimensional equation $x_3(u_3)$. With the obtained value of $u_3$, the equation
$x_2(u_2,u_3)$ becomes a one dimensional function of $u_2$ which can be
inverted. Last with the obtained $(u_2,u_3)$, one can invert $x_1(u_1,u_2,u_3)$
to obtain $u_1$.

\header{Product Model} While the coordinate transform described above is very
general, it is only useful if the multiple integrals in \eqnref{eq:l_from_u_app}
can be performed efficiently. Otherwise, it is as or more costly than the actual
integral of the perturbation series coefficient that we wish to compute. 

The product model function
\begin{equation}
\tag{\ref{eq:warper1d}$'$}\label{eq:warper1d_app}
p_n(\vec{u}) = \prod_{i = 1}^{n} h^{(i)}\bigl( u_{i-1} - u_{i} \bigr),
\end{equation}
is particularly efficient. It is simpler to view this as a composition of two
transforms. First, we change variables $v_i = u_{i-1} - u_{i}$, which has
Jacobian $\left\vert {\partial \vec{u}}/{\partial \vec{v}}\right\vert = 1$.
Second, in the $v_i$ variables, the coordinate transform
\eqnref{eq:l_from_u_app} separates entirely, so that each $x_m$ only depends on
a single variable $v_m$:
\begin{equation}
   \label{eq:x_of_v_simple}
x_{m}(v_{m})  =  \frac{\int_0^{v_m}  \ud \bar{v}_{m} \,\,  h^{(m)}(\bar{v}_m)}{\int_0^{\infty}  \ud \bar{v}_{m} \,\, h^{(m)}(\bar{v}_m)}
\end{equation}
These one-dimensional integrals can be integrated quickly and precisely
analytically or using quadrature algorithms. We then invert $x_{m}(v_{m})$
numerically to find the coordinate transform $v_m(x_m)$. In practice, it is
possible to completely pre-compute these integrals on a fine mesh for fast
evaluation during calculation.

We note that the product form \eqnref{eq:warper1d_app} is known for importance
sampling in Monte Carlo applications, e.g.~as part of the VEGAS algorithm
\cite{Lepage_1978, Lepage_1980}. The choice of coordinate system $v_i$ as
compared to $u_i$ affects the quality of model function and is an important
physical consideration.

\header{MPS Model} The model function \eqnref{eq:warperMPS} can also be
efficiently computed using the above algorithm and standard MPS techniques
\cite{Schollwoeck_1008}. Unlike the product model, the integrations for
different $v_i$ have to be performed in sequence, with a matrix-vector
multiplication at each step.

\section{Explicit expressions of the integrands}
\label{app:manybody}

In this appendix, we will describe the Anderson impurity model as well as the
perturbation expansion formalism of our calculations in more detail.

\subsection{Model}
The Hamiltonian of the Anderson impurity model is $H = H_0 + H_{\rm
int}\theta(t)$, with the non-interacting term:
\begin{equation}
    H_0 = \sum_{i,\sigma} \left( \gamma_i c^{\dagger}_{i,\sigma} c^{\phantom{\dagger}}_{i+1,\sigma} + H.c. \right)+ E_d  \sum_\sigma c^{\dagger}_{0\sigma} c^{\phantom{\dagger}}_{0\sigma}.
\end{equation}
The hopping parameters are all $\gamma_i = D/2$, except at the impurity ($i=0$)
where $\gamma_0 = \gamma_{-1} = \gamma$. The interaction term is 
\begin{equation}
H_{\rm int} = U (c^{\dagger}_{0\uparrow} c^{\phantom{\dagger}}_{0\uparrow}
- \alpha) (c^{\dagger}_{0\downarrow} c^{\phantom{\dagger}}_{0\downarrow} - \alpha), 
\end{equation}
where $\alpha$ is a quadratic shift to the perturbation. This shift means that
the $U$ expansion is performed about a different starting point, and is commonly
used to improve perturbation series convergence
\cite{Profumo_1504,Rubtsov2004,Wu_1608}. Note that the energy of a single
electron localized on the impurity is $\vare_d = E_d - \alpha U$. A symmetric
voltage bias $V$ is applied between the two leads. As is standard in the Keldysh
formalism, $H_{\mathrm{int}}$ is turned on at time $t=0$ and observables are
computed after a large time $t$ when the stationary regime has been reached.

By integrating out the leads, their effect on the dot is represented by a
retarded hybridization function $\Delta(\omega)$. The non-interacting dot
retarded Green function is $ g^R(\omega) \equiv 1 / (\omega - E_d -
\Delta(\omega)) $. By symmetry, all Green functions are the same for spin up and
down.

The density of states of the leads is semi-circular with half-bandwidth $D$. An
important parameter of the non-interacting model is the tunneling rate from the
impurity to the leads at the (equilibrium) Fermi level $\Gamma = 4\gamma^2 / D$.
In terms of $D$ and $\Gamma$, the hybridization function is
\begin{equation}
    \Delta(\omega) = \frac{\Gamma}{D} \times
    \begin{cases}
        \left( \omega + \sqrt{\omega^2 - D^2} \right) & \text{for $\omega < -D$} \\
        \left( \omega - i \sqrt{D^2 - \omega^2} \right) & \text{for $-D < \omega < D$} \\
        \left( \omega - \sqrt{\omega^2 - D^2} \right) & \text{for $\omega > D$}
    \end{cases}
\end{equation}

The Bethe ansatz provides results only in the universal regime, where $D
\rightarrow +\infty$ with $\Gamma$ fixed. In this limit, the density of states
becomes independent of energy (flat band). The hybridization function is simply
$\Delta(\omega) = -i\Gamma$. When comparing results to the Bethe ansatz, we we
will always work in this regime.

\subsection{Expressions for the integrands of series expansions}
To obtain the number of electrons $Q$ or the current $I$, we compute a
perturbation series in $U$ for the equal-time lesser Green function $G^<_{0i}(t,
t) = -i\langle c^\dagger_{0\uparrow}(t) c^{\phantom{\dagger}}_{i\uparrow}(t)
\rangle$, where the creation and annihilation operators are in the Heisenberg
picture. We take $i=0$ (on-site) to obtain $Q$, and $i=1$ (dot-lead) to obtain
$I$ \cite{Meir_1992}.

For $\alpha = 0$, this series can be written compactly using the ``Wick
determinant" notation introduced in Ref.~\cite{Bertrand_1903_kernel}:
\begin{multline}
    \label{eq:integral_gf}
    G^<_{0i}(t, t) = \sum_{n\ge 0} \frac{i^nU^n}{n!} \int_0^t \prod_{k=1}^n du_k\\
    \times \sum_{a_1, \ldots, a_n} (-1)^{\sum a_k} \lists{(0, t, 0), U_1, \ldots, U_n}{(i, t, 1), U_1, \ldots, U_n}
    \lists{U_1, \ldots, U_n}{U_1, \ldots, U_n},
\end{multline}
where $a_k \in \{0, 1\}$ are Keldysh indices, and $U_k = (0, u_k, a_k)$
represents a point on the Keldysh contour composed of a site index (here $0$ for
the impurity), a time $u_k$ and a Keldysh index $a_k$. The Wick determinant
$\lists{\ldots}{}$ is defined, for $A_1, \ldots, A_m$ and $B_1, \ldots, B_m$ any
set of points on the Keldysh contour, in the case $\alpha = 0$, the Wick
determinant 
\begin{equation}
    \lists{A_1, \ldots, A_m}{B_1, \ldots, B_m} =
    \begin{vmatrix}
        {g}(A_1, B_1) & \hdots & {g}(A_1, B_m) \\
        \vdots & \ddots & \vdots \\
        {g}(A_m, B_1) &  \hdots & {g}(A_m, B_m)
    \end{vmatrix},
\end{equation}
where
\begin{equation}
    g[(x, u, a), (x', u', a')] =
    \begin{pmatrix}
        g^T_{xx'}(t, t') & g^<_{xx'}(t, t') \\
        g^>_{xx'}(t, t') & g^{\bar T}_{xx'}(t, t')
    \end{pmatrix}_{aa'},
\end{equation}
is the non-interacting one-particle Keldysh Green function. Here $g^T$, $g^{\bar
T}$, $g^<$ and $g^>$ are respectively the time-ordered, anti-time-ordered,
lesser and greater Green functions. In \eqnref{eq:integral_gf}, the determinant
on the left is from spin up operators, while the one on the right is from spin
down operators. Nevertheless, by spin symmetry their elements share the same
Green functions $g$.

The case $\alpha \ne 0$ is similar, but the diagonal terms of the Wick
determinants in \eqnref{eq:integral_gf} must be shifted by $-i\alpha$, except
the one connecting to the measurement point (involving the Green function at
time $t$) \cite{Profumo_1504}.

Provided that the non-interacting Green function $g$ is known as a function of
time, \eqnref{eq:integral_gf} explicitly defines the integrand that we refer to
in the main text of this article. We compute the time domain Green functions by
Fourier transform of the Green functions $g^<(\omega)$ and $g^>(\omega)$. These
can be derived in the Schwinger--Keldysh formalism \cite{Meir_1992}:
\begin{align}
    \label{eq:lesser_greater_freq}
    g_{00}^<(\omega) = & 2 i |g^R(\omega)|^2 \Im[\Delta(\omega)] \\
                       & \times \left[ n_{\rm F}\left(\omega - \frac{V}{2}\right) + n_{\rm F}\left(\omega + \frac{V}{2}\right) \right] \nonumber\\
    g_{00}^>(\omega) = & g_{00}^<(\omega) - 4 i |g^R(\omega)|^2 \Im[\Delta(\omega)] \\
    \gamma g_{01}^<(\omega) = & \frac{\Delta(\omega)}{2}g^<_{00}(\omega) - 2i\; n_{\rm F}\!\left(\omega + \frac{V}{2}\right) \Im[\Delta(\omega)]g^R(\omega)^*
\end{align}
Here $n_\mathrm{F}(\omega)$ is the Fermi function. Note that the function
$g_{0i}^>$ is not used in \eqnref{eq:integral_gf}.

\begin{table*}
\begin{tabular}{c | l | l }
{$n$} & \multicolumn{2}{c}{Occupation Number $Q^\mathrm{Bethe}_n$} \\
\hline
  $0$ & $ 1 $&$ \phantom{-}1.000000000000000000000000 $\\
  $1$ & $ -\frac{1}{\pi } $&$ -0.3183098861837906715377675 $\\
  $2$ & $ \frac{1}{\pi ^2} $&$ \phantom{-}0.1013211836423377714438795 $\\
  $3$ & $ \frac{\pi ^2-9}{3 \pi ^3} $&$ \phantom{-}0.009348692094998422959323018
  $\\
  $4$ & $ \frac{90-11 \pi ^2}{6 \pi ^4} $&$ -0.03176576952402088647648727 $\\
  $5$ & $ -\frac{9450-1175 \pi ^2+18 \pi ^4}{90 \pi ^5} $&$
  \phantom{-}0.01428453978718527764087931 $\\
  $6$ & $ \frac{945}{\pi ^6}-\frac{3185}{27 \pi ^4}+\frac{137}{60 \pi ^2} $&$
  \phantom{-}0.003296943760155087659102276 $\\
  $7$ & $ -\frac{10395}{\pi ^7}+\frac{11690}{9 \pi ^5}-\frac{2653}{100 \pi
  ^3}+\frac{1}{7 \pi } $&$ -0.007417285598070720865765088 $\\
  $8$ & $ \frac{135135}{\pi ^8}-\frac{912065}{54 \pi ^6}+\frac{524881}{1500 \pi
  ^4}-\frac{363}{140 \pi ^2} $&$  \phantom{-}0.003074980697930490893903131 $\\
  $9$ & $ -\frac{2027025}{\pi ^9}+\frac{41046005}{162 \pi
  ^7}-\frac{118576073}{22500 \pi ^5}+\frac{623873}{14700 \pi ^3}-\frac{1}{9
  \pi } $&$ \phantom{-}0.001257173321207511996877863 $\\
 $10$ & $ \frac{34459425}{\pi ^{10}}-\frac{348898550}{81 \pi
 ^8}+\frac{5045327287}{56250 \pi ^6}-\frac{761337511}{1029000 \pi
 ^4}+\frac{7129}{2520 \pi ^2} $&$ -0.002096420663210080104411625 $\\
 $11$ & $ -\frac{654729075}{\pi ^{11}}+\frac{19887342475}{243 \pi
 ^9}-\frac{287681226833}{168750 \pi   ^7}+\frac{306048256943}{21609000 \pi
 ^5}-\frac{4785253}{79380 \pi ^3}+\frac{1}{11 \pi } $&$
 \phantom{-}0.0007382700406215484131765644 $\\
 $12$ & $ \frac{13749310575}{\pi ^{12}}-\frac{2505809831525}{1458 \pi
 ^{10}}+\frac{90628717412233}{2531250 \pi   ^8}-\frac{84543422632097}{283618125
 \pi ^6}+\frac{131145705977}{100018800 \pi ^4}-\frac{83711}{27720 \pi ^2} $&$
 \phantom{-}0.0004844375173694184755334739 $\\
\end{tabular}
\caption{Perturbation coefficients for the occupation number from the Bethe Ansatz.}
\label{tab:BA_Qcoeff}
\end{table*}

\subsection{Precision calculation of $g^<(t)$ and $g^>(t)$}

Care has to be taken when performing the Fourier transform to obtain $g^<$ and
$g^>$ in the time domain. As integration methods get increasingly precise, the
accuracy of the integrand becomes more important. In order to provide benchmarks
with relative error of $\sim 10^{-6}$ at order $n \sim 10$, and to rule out any
bias due to inexact integrands, we need to refine the calculation of $g(t)$.  In
particular, using a Fast Fourier Transform (FFT) algorithm produces an error
which decreases too slowly with the number of samples for functions with sharp
features or power law tails, such as the ones we encounter here.

One approach to high precision is to compute the Fourier transform using
adaptive quadrature methods. This is precise enough if a system has a finite
bandwidth and the integrand is proportional to the density of states of the
leads and therefore has bounded support.  In general, however, the integrand
decays slowly and oscillates at high frequencies, which renders direct
integration methods inaccurate. Alternatively, when the tails are dominated by
simple poles, it is possible to separate them out analytically and perform the
finite remainder using a FFT. In our case, however, the tails are dominated by
the Fermi functions, which have an essential singularities at $\vert\omega\vert
= \infty$, and we must resort to other methods.

To circumvent this problem, we deform the integration path in the $\omega$
complex plane to find a more favorable integrand, and apply an adaptive
quadrature method. We show that a path can be found for a generic class of
problems which improves the decay rate and eliminates oscillations near
infinity. We consider the general case of finite temperature, and denote the
inverse temperature by $\beta$. We will work out the case for $g^<$, but $g^>$
can be treated equally by first applying a change of variable $\omega
\rightarrow -\omega$.
Specifically, at time $t$, the Fourier Transforms we are interested in can
always be decomposed in a sum of integrals of the form:
\begin{equation}
    \label{eq:ft_integral}
    \int d\omega' \; \zeta(\omega') n_{\rm F}(\omega') e^{-i\omega' t}
\end{equation}
with $\omega' = \omega \pm V/2$. The function $\zeta$ depends on $g^R$ and
$\Delta$. Its exact form does not affect the choice of a new path, as long as it
has no singularity at $|\omega'| = \infty$. We will further assume that one can
bound the complex singularities of $\zeta$ and $n_{\rm F}$ inside a vertical
band $\omega^- < \Re[\omega'] < \omega^+$.

For $t \ne 0$, we compute the integral \eqnref{eq:ft_integral} along a new path
parametrized by $x$ and defined as:
\begin{equation}
    \omega'(x) =
    \begin{cases}
        \omega^- - (x - \omega^-) i t, & \text{for $x < \omega^-$}, \\
        x, &\text{for $\omega^- < x < \omega^+$}, \\
        \omega^+ + (x - \omega^+) (\beta -it) , &\text{for $x > \omega^+$}.
    \end{cases}
\end{equation}
The new path is made of three pieces, joined together at $\omega' = \omega^\pm$.
The central one is simply a segment of the real axis, left unchanged to prevent
crossing singularities. The other two are straight lines at an angle with the
real axis, which have been chosen so that the integrand becomes asymptotically
proportional to a decaying, oscillation-free, exponential. The points
$\omega^\pm$ can be moved away from one another, in particular to avoid the $x <
\omega^-$ piece being too close to singularities. For $\beta = +\infty$, the
integrand is zero on the half-plane $\Re[\omega'] > 0$, hence the $x > \omega^+$
piece of the path can be ignored. In the case $t=0$, the integrand is simply
$g_{0i}^<(\omega)$, which is free of oscillations, and deforming the integration
path would make some appear. Hence the integration path  is left untouched in
this case.

This technique is easily generalized to more complex impurity systems. However,
it relies on an analytical continuation and knowledge of singularities. This may
not be easily accessible for numerically computed Green functions.

\section{Bethe Ansatz Comparison}
\label{app:BA}

Here we briefly discuss the Bethe ansatz solution for the Anderson impurity
model and how specifically we extract the coefficients $Q_n^{\mathrm{Bethe}}$
from the general solution (see \cite{Tsvelick_1983_R,Okiji_1984} and references
therein).  We are interested in the case were $\vare_d = 0$ and we then perform
a perturbative expansion in $U$. This always corresponds to the so-called
symmetric limit $U/2 + \vare_d \ll \sqrt{U\Gamma}$. 

We use the results of \cite{Wiegmann_1983_a}. In the symmetric limit, the
occupation number $Q$ on the quantum dot is given by a series 
\begin{multline}
\label{eq:BA_n_series}
Q = 1 - \sum_{n=0}^{\infty} \frac{\sqrt{2}}{\pi} \frac{(-1)^n}{(2n+1)} G^{(+)}[i\pi(2n+1)]   \\ \cdot\int_{-\infty}^{\infty} d k \left\{ \Delta(k) e^{-\pi (2n+1) [g(k) - \Lambda]} \right\}.
\end{multline}
Here:
\begin{align}
\Delta(k) &= \frac{\Gamma}{\pi} \cdot \frac{1}{(k-\vare_d)^2 + \Gamma^2}, \\
g(k) &= \frac{(k - \vare_d - U / 2)^2}{2 U \Gamma}, \\
G^{(+)}(\omega) &= \frac{\sqrt{2\pi}}{\Gamma\left(\tfrac{1}{2} - \frac{1}{2\pi} i \omega \right)} \left(\frac{-i\omega + 0}{2\pi e}\right)^{-i\omega / 2\pi}
\end{align}
and $\Lambda$ is the energy cutoff. All other quantities are the same as in
Appendix~\ref{app:manybody}. The cutoff $\Lambda$ is given implicitly by the
series:
\begin{align}
\label{eq:BA_cutoff}
\frac{U/2 + \vare_d}{\sqrt{U \Gamma}} = \frac{1}{2}\cdot \frac{2}{\pi} \sum_{n=0}^\infty \frac{(-1)^n G^{(+)}[i \pi (2 n + 1) ] e^{\pi \Lambda (2n +1)} }{(2n + 1)^{3/2}}
\end{align}

We emphasize that \eqnref{eq:BA_cutoff} differs from the expression in
\cite{Wiegmann_1983_a} by a factor of $1/2$. The constraint on $\Lambda$ arises
from imposing number conservation on the distribution function, which we take to
be of the form used by \cite{Okiji_1984}.

We now set $\vare_d = 0$, so that our expansion parameter is $\sqrt{U/\Gamma}
\ll 1$. To extract the coefficients $Q_n$ we use the following procedure. First,
write \eqnref{eq:BA_cutoff} as a polynomial in $x = e^{\Lambda \pi}$ up to order
$N_{\mathrm{cutoff}} = 30$. Second, perform polynomial inversion to find $x$ as
a function of $\sqrt{U/\Gamma}$, repeatedly using the smallness of
$\sqrt{U/\Gamma}$. Third, evaluate the $k$ integral in \eqnref{eq:BA_n_series}
analytically in an asymptotic expansion in $\sqrt{U/\Gamma}$ up to order
$N_{\mathrm{cutoff}} $, (see e.g.~\cite{Horvatic_1985}). Fourth, substitute the
expansion of $x$, collecting terms of the same order in $U/\Gamma$. The final
result is a power series expansion in $U/\Gamma$, with analytic coefficients.
These are evaluated numerically with high precision arithmetic and shown in
Table~\ref{tab:BA_Qcoeff}. For orders $n=0 - 5$, the analytic expressions match
the results calculated explicitly in perturbation theory in
\cite{Horvatic_1985}.

\section{Error Calculation in Quasi-Monte Carlo}
\label{app:error_calc}

If we have reached the asymptotic regime for a single sequence $Q(N)$, we can
estimate an error by fitting the approach to convergence. This is similar to
error estimation for traditional quadrature. 

\begin{figure}[t]
    \centering
    \includegraphics[width=\columnwidth]{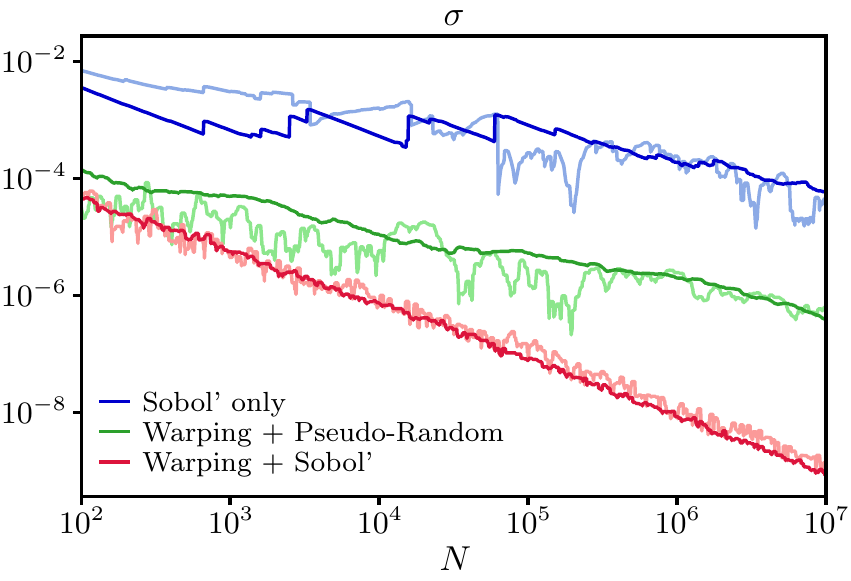}
    \caption{
	\label{fig:error-calc}
	Absolute error of occupation number $Q_6(N)$, with curves matching three
	cases of Fig.~\ref{fig:Qn-conv-warpers}. Here we perform calculations with
	$K=25$ shifted sequences. The pale curves show the convergence of the
	average over these runs to the Bethe ansatz result. The solid curves show
	the corresponding uncertainty estimates computed from the standard error of
	the mean over the shifted sequence values. We see that this procedure gives
	good estimates of the error.}
\end{figure}

A more robust estimate of the error can be achieved by reintroducing a random
component to the method -- so-called Randomized Quasi-Monte Carlo
\cite{Dick_2013,Nuyens_1308,Dick_Pillichshammer_2010,Lecuyer_2018}. Here we
repeat the calculations with $K$ `randomly shifted' sequences giving a
distribution of values $Q_k$, from which we obtain an estimate of the mean and
error as for conventional Monte Carlo. Typically one chooses only a moderate
number of $K \sim 10 - 100$ as it is advantageous, for fixed computational time,
to maximize $N$.  
In Fig.~\ref{fig:error-calc}, we show this approach in practice for different
methods of Fig.~\ref{fig:Qn-conv-warpers}.

\section{Details on the current $I$ calculation}
\label{app:current_details}

This appendix gives details of the calculations made to obtain
Fig.~\ref{fig:current}.

In this application, each lead has centered, semi-circular density of states
with half-bandwidth $D=2$. The coupling to the leads is chosen so that $\Gamma =
0.1$. A symmetric bias voltage $V$ is applied between the two leads. For each
parameter set ($\vare_d, V, U$), $E_d$ and $\alpha$ are chosen so that the first
order (Hartree term) of the perturbation series vanishes at $U = 7\Gamma$, to
improve its convergence radius. The series, defined in
Appendix~\ref{app:manybody}, is computed at a time $t=10 / \Gamma$ after
switching the interaction on.

Due to the finite bandwidth $D$, the integrand decays polynomially at large
times, but exponentially at intermediate times. We used an exponential warping
with $\tau = 1.5 / \Gamma$, which was enough to capture the general shape of the
integrand up to the observation time $t$.

\begin{figure}[t]
    \centering
    \includegraphics[width=\columnwidth]{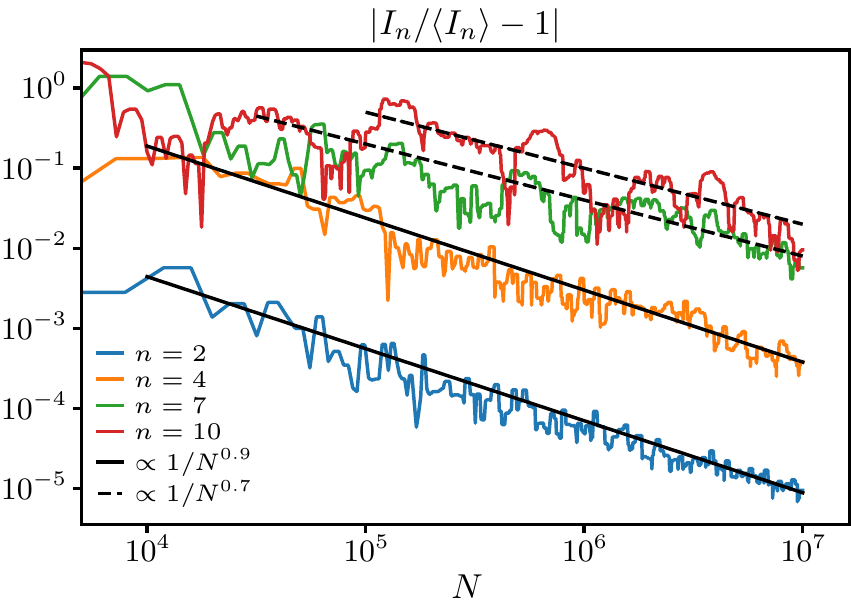}
    \caption{
	\label{fig:convergence-current}
    Convergence of the current $I_n$ with number of samples $N$ at different
    orders $n$. This data corresponds to a typical point of
    Fig.~\ref{fig:current}, away from particle-hole symmetry ($\vare_d/U \approx
    -0.12$, $\alpha \approx 0.27$, $U = 6\Gamma$). At low order (blue and orange
    lines) the relative error scales as $1/N^{0.9}$ (black plain line). At
    larger orders ($n=7$ and $10$, green and red lines), the convergence slows
    down and scales only as $1/N^{0.7}$ (dashed black line). Finally at order
    $n=10$, the final scaling starts at a larger number of function evaluations
    $N$ than for lower orders. For visibility, the data has been smoothed as
    described in Appendix~\ref{app:current_details} (see main text).}
\end{figure}

Figure \ref{fig:convergence-current} shows the convergence of the calculation
for a system away from particle-hole symmetry ($E_d \approx 0.9\Gamma$, $\alpha
\approx 0.27$, which in Fig.~\ref{fig:current} corresponds to the point
$\vare_d/U \approx -0.12$ and $U = 6\Gamma$). We observe a scaling slightly
below $1/N$, which deteriorates with increasing order. Also, at large orders,
the final scaling regime is reached at a later $N$.

Note that the data of Figs.~\ref{fig:Qn-conv-orders}, \ref{fig:Qn-conv-warpers}
and \ref{fig:convergence-current} have been smoothed for visibility: for
$N>100$, we show the maximum of the error in a moving window around $N$ of fixed
size in log-space ($5\%$ of $N$). This smoothing generates an upper bound of the
error.

\begin{figure}[t]
    \centering
    \includegraphics[width=\columnwidth]{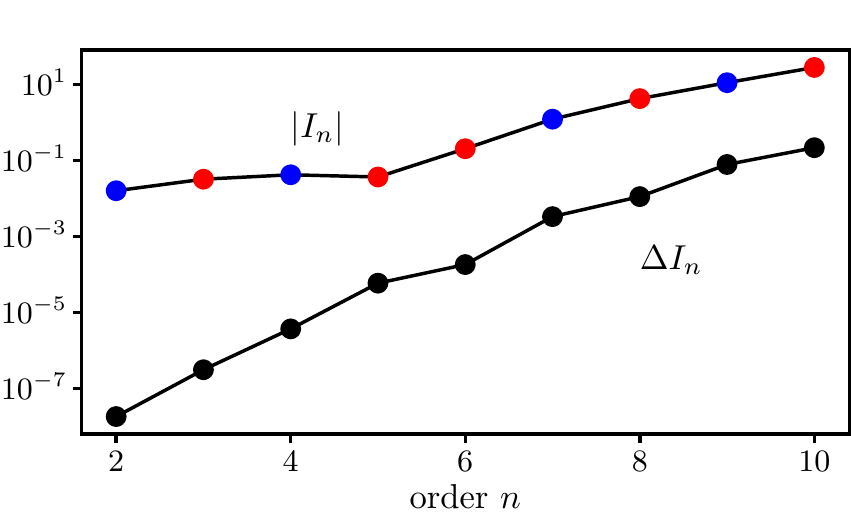}
    \caption{
	\label{fig:series-current}
    Perturbation series of the current for the same parameters as in
    Fig.~\ref{fig:convergence-current}. The upper line (colored symbols) gives
    the absolute values of the series coefficients $\vert I_n \vert$: positive
    coefficients are depicted in blue, and negative ones in red. The lower line
    (black symbols) is the estimated absolute error $\Delta I_n$.  For all
    orders calculated, up to $n=10$, we obtained at least two digits of
    accuracy.}
\end{figure}

For each integration, digitally shifted \cite{Kuo_1606} generators are used to
produce 10 different Sobol' sequences from which we take the average to obtain
the final result $\langle I_n \rangle$. The error is estimated by taking the
standard deviation of the 10 results and dividing by $\sqrt{10}$. An example of
a series computed with its estimated error is shown in
Fig.~\ref{fig:series-current}. The error (black dots) on the coefficients
(colored dots) is low enough so that only the truncation of the series limits
the resummation accuracy.

The convergence radii of the series are about 3--4 $\Gamma$, so the series are
resummed to obtain answers at larger interaction $U$. We use the conformal
transform technique of Ref.~\cite{Bertrand_1903_series} with the so-called
parabola transform $W = -\tan^2(\sqrt{U/p})$. Here $p$ is a real negative
parameter optimized for each series. Depending on their analytical structure,
the series $I(U)$ or its inverse $1/I(U)$ is resummed, whichever gives smaller
error. The integration error is propagated through the resummation process and
added to the truncation error, the latter being estimated from the convergence
radius of the resummed series. Unlike in Ref.~\cite{Bertrand_1903_series}, no
Bayesian inference is used.

At large $\vare_d$, the integrands become more difficult to integrate, but at
the same time the summation of the series requires less orders for the same
precision. When $\vare_d / U > 0.5$ (or $\vare_d / U < -1.5$ by symmetry) we
only computed and summed the series up to order $n=5$.

Finally, it is worth noting that small bias $V$ setups reach a stationary regime
after a longer time. Hence performing the integration away from the linear
response regime, at large $V$, is actually less expensive and less prone to sign
problem. Physically, the voltage reduces the coherence of the system, which
explains the reduction of almost exact cancellations in the numerical
integration.


\section{Construction of the 1D  Model Function}
\label{app:model_function_details}

In this appendix, we explain how the $h^{(i)}$ functions are constructed in the
model function defined in the main text,  
\begin{equation}
\tag{\ref{eq:warper1d}$'$}\label{eq:warper1d_appE}
p_n(\vec{u}) = \prod_{i = 1}^{n} h^{(i)}\bigl( u_{i-1} - u_{i} \bigr).
\end{equation}
for $0 < u_n < u_{n-1} < \ldots <u_2 <u_1$ and $u_0 = t$ is defined to be the
measurement time.

We are going to use successive changes of variable in this section:
\begin{equation}
\vec u \rightarrow \vec v \rightarrow \vec w \rightarrow \vec x
\end{equation}
First, we change to the $v_i$ variables defined as $v_i \equiv u_{i-1} - u_{i} >
0$ which are natural since the integrand only depends on time differences. The
corresponding Jacobian is one. Note that the integration on $\vec v$ is
performed on $[0,\infty)^n$. When going back to $\vec u$ space, this generates
extra points that are not in the original $\vec u$ domain. The value of the
integrand for these points is simply zero so that they induce no extra
computational cost. 

\subsection{Exponential form for $h^{(i)}$}

\begin{figure}[t]
    \centering
    \includegraphics[width=\columnwidth]{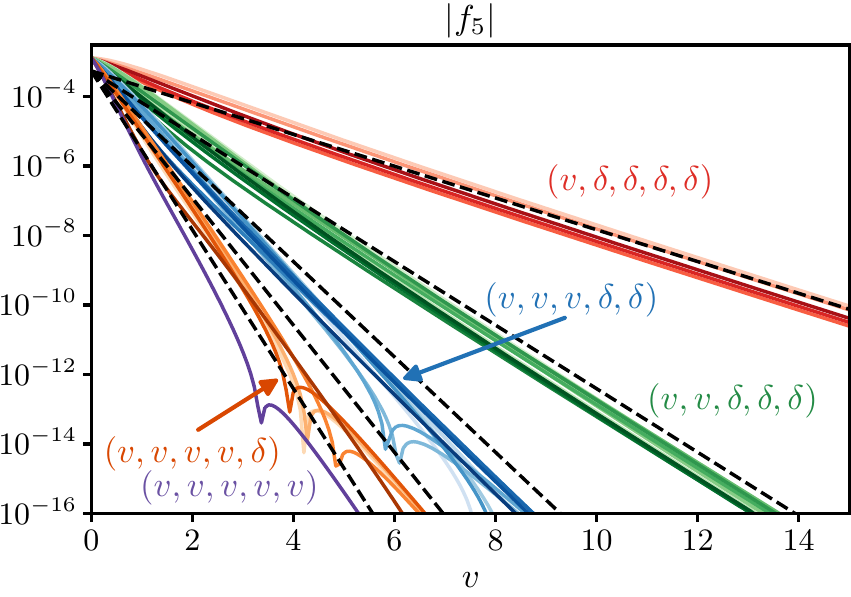}
    \caption{
    \label{fig:f-directions}
    Comparison between the absolute value of the order 5 integrand $|f_5(\vec
    v)|$ in \eqnref{eq:def_int} (colored lines) and the model function $p_5(\vec
    v)$ of \eqnref{eq:warper1d_appE} (black dashed lines, $\tau=0.95$) along
    various directions in $\vec v$-space. Each color corresponds to $\vec v =
    (v, \ldots, \delta)$ as indicated in the figure. Lines of the same color
    correspond to different permutations within a given direction (see main
    text). The parameters are the same as in Fig.~\ref{fig:Qn-comparison}.}
\end{figure}

Our second change of variable $ \vec v \rightarrow \vec w$ will be based on a
model function with a simple analytic form for $h^{(i)}$, designed to correctly
describe the asymptotics of the integrand. This asymptotic region becomes
increasingly important at large order $n$, especially in the long time limit
$t\rightarrow \infty$. We choose a simple form, independent of $i$
\begin{equation}
   \label{eq:hi_expo}
   h^{(i)}_{\rm exp} (v) = e^{- v/\tau}, \qquad \forall i.
\end{equation}
This choice is motivated by a direct study of the integrand. In
Fig.~\ref{fig:f-directions}, we plot the absolute value of the integrand
$|f_5(v_1,v_2,v_3,v_4,v_5)|$ of $Q_5$ with the parameters of our benchmark along
various directions of the $5$ dimensional $\vec v$-space. The different colors
correspond to different directions: red corresponds to
$(v,\delta,\delta,\delta,\delta)$ where $\delta$ is fixed to $\delta=0.5$ and
$v$ is varied. The 5 different curves correspond to different permutations of
the $v$ with respect to the $\delta$. Green curves correspond to the 10
different permutations of $(v,v,\delta,\delta,\delta)$ and so on. The dashed
line corresponds to \eqnref{eq:hi_expo} with $\tau=0.95$. Remarkably, such a
simple ansatz with a \emph{single} parameter, already captures the integrand
asymptotics well, in various directions in 5 dimensions.

For a fixed number of integrand evaluations $N$, the error made in the
calculation of $Q_n$ is very sensitive to the choice of the parameter $\tau$ as
shown in Fig.~\ref{fig:f-tau}. For $N=10^6$, Fig.~\ref{fig:f-tau} shows the
relative error as a function of $\tau$ for various orders $n=5,6,7$ and $10$.
The error possesses a sharp minimum around $\tau=0.85$ (note the log scale). If
$\tau$ is too small, we under-sample the tails of the function leading to
potentially incorrect results.  If $\tau$ is too large, the calculation is
correct, but less efficient since the sampling puts a lot of points in regions
which contribute little to the result. We will now see how to make the
computation more robust regarding the choice of $\tau$.

\subsection{Learning  $h^{(i)}$ from the integrand}
\label{app:optim_hi}

\begin{figure*}[t]
    \centering
    \includegraphics[width=\textwidth]{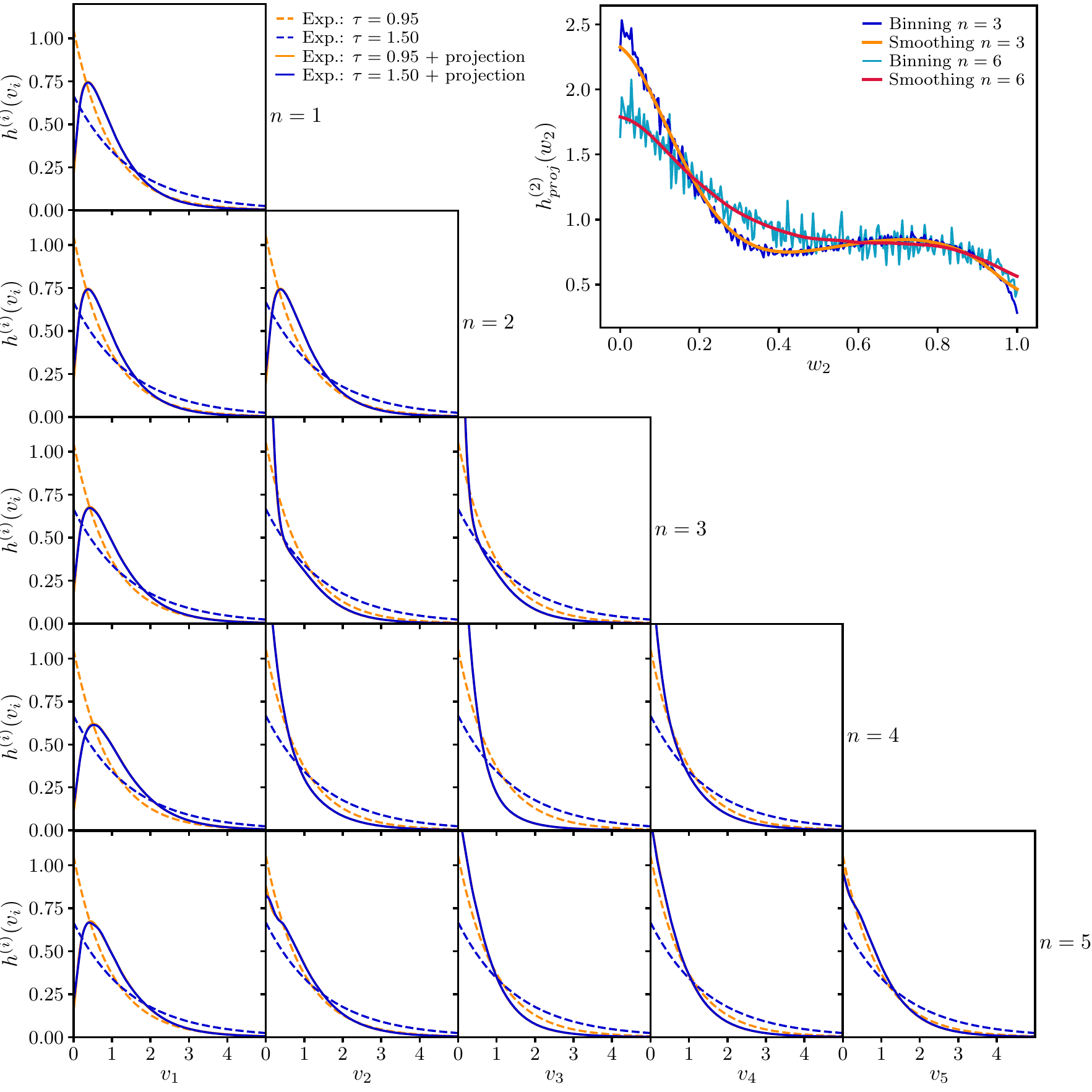}
    \caption{\label{fig:optim_hi}
        Examples of functions $h^{(i)}(v_i)$ for various $n$ and $1 \leq i \leq
	n$: exponential form (dashed lines) and after projection  (plain lines, on
	top of each other). We used $N_b = 250$ bins, $M=20000$ samples and $\lambda
	= 0.05$; see text. An example of binned values and smoothing for
	$h^{(2)}_{\rm proj}$ for order $n=3,6$ is shown in the upper right inset.   
        The parameters are the same as in Fig.~\ref{fig:Qn-comparison}.}
\end{figure*}

In this section, we make an additional change of variable $\vec w \rightarrow
\vec x$. To approximate the new integrand $\tilde f_n(\vec w)$ in the $\vec w$
variable, we search for a new model function $\tilde p_n(\vec w)$, again having
the form:
\begin{equation}
\tilde p_n(\vec{w}) = \prod_{i = 1}^{n} \tilde h^{(i)}\bigl( w_{i} \bigr).
\end{equation}
If we assume that the integrand $\tilde f_n(\vec{w})$ is well approximated by
such a simple form, we estimate the functions $\tilde h^{(i)}$ by projecting in
each dimension 
\begin{equation}
   \tilde h^{(i)}_{\rm proj}(y) \equiv \int \ud^n\vec{w} \tilde f_n(\vec{w})  \delta(w_i - y).
\end{equation}
In practice, we calculate $M$ values of the integrand $\tilde f(\vec w_\alpha)$,
$\alpha\in 1...M$ generated by sampling the $\vec w$ space using the Sobol'
sequence, and bin them in each dimension with $N_b$ bins. The function that we
obtain is rather noisy due to the binning, so in a second step, we use a
Gaussian kernel smoothing
\begin{equation}
\tilde h^{(i)}_{\rm proj}(w) \rightarrow \frac{ 
\sum_{l=1}^{N_b}  K_\lambda \left(w,\frac{l}{N_b}\right) \tilde h^{(i)}_{\rm proj}\left(\frac{l}{N_b}\right)
}{
\sum_{l=1}^{N_b}  K_\lambda \left(w,\frac{l}{N_b}\right)
}
\end{equation}
where 
\begin{equation}
K_\lambda(w, w') = e^{{-(w-w')^2/\lambda^2}}.
\end{equation}
The upper right plot of Fig.~\ref{fig:optim_hi} shows an example of the
smoothing procedure.  

\begin{figure}[t]
    \centering
    \includegraphics[width=\columnwidth]{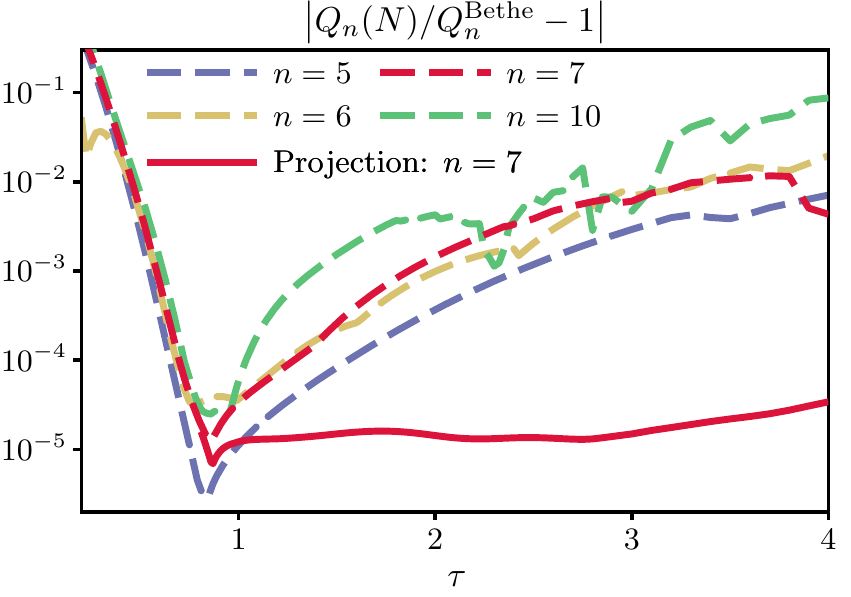}
    \caption{\label{fig:f-tau}
    Exact relative error of $Q_n$ computed with $N=10^6$ points as a function of
    the model function parameter $\tau$. The dashed lines are for a model
    defined by a single exponential function $h^{(i)}(v_i) = \exp(-v_i/\tau)$,
    at different orders. The red solid curve is the new error after optimization
    of $h^{(i)}$ by projection (see section~\ref{app:optim_hi}), at order $n=7$.
    The parameters are the same as in Fig.~\ref{fig:Qn-comparison}.}
\end{figure}

The change of variable based on  $\tilde h^{(i)}_{\rm proj}$ was used to compute
the continuous line in Fig.~\ref{fig:f-tau} for order $n=7$ ($N_b = 100, M =
10^5, \lambda = 0.05$).  The error is improved at the optimum point $\tau =0.95$
only by approximately a factor 2.  However, it remains largely independent of
the value $\tau$ selected for the $\vec v \rightarrow \vec w$ change of
variable, showing that the method has become much more robust.  The projection
automatically fits the exponential tails, without manual adjustment (as long as
$\tau$ is not too small, to avoid under-sampling as explained above).
Note that in Fig.~\ref{fig:f-tau}, only $M=10^5$ values of the integrand were
used for the learning step. This is 1\% of the total number of function calls,
hence negligible. If $\tau$ is increased, the sampling of the integrand
decreases in quality, as the tails are over-sampled. Correcting this in the
learning step becomes increasingly more demanding. Therefore the change of
variable $\vec v \rightarrow \vec w$ with a good initial guess for $\tau$ is
essential to the success of the projection.

Because we only used model functions of the form \eqnref{eq:warper1d_appE},
according to \eqnref{eq:x_of_v_simple}, $x_m$ only depends on $w_m$, which
itself only depends on $v_m$. In this special situation, the change of variable
$\vec v \rightarrow \vec x$ can be represented by a model function of the form
\eqnref{eq:warper1d_appE}, and compared to the simpler $\vec v \rightarrow \vec
w$. As a result, Fig.~\ref{fig:optim_hi} shows $h^{(i)}_{\rm proj}(v_i)$ along
with the initial exponential guess, for all $i$ and $n\leq 5$, and for two
values of $\tau=0.95$ and $\tau = 1.5$. $h^{(i)}_{\rm proj}$ computed from two
different initial guess for $\tau$ are indistinguishable (plain lines, on top of
each other), showing that the result is independent of the initial choice of
$\tau$. At small $v$, they are quite different from a pure exponential. Finally,
we note that the projected model functions vary only slightly with $n$ for most
values of $i$. This could be turned into an advantage by reusing the $h$
functions from a lower dimension to a higher one in future developments.

\subsection{Comparison to VEGAS algorithm}

The form of model function \eqnref{eq:warper1d_app} and the projection procedure
described above are related to the VEGAS algorithm \cite{Lepage_1978,
Lepage_1980}. However, simply applying this algorithm to perform the
perturbation theory integrals produces poor results. Here we describe the
essential differences between VEGAS and our approach.

The standard VEGAS algorithm uses random numbers to sample the integrand. It
generates a weight function by projecting samples onto coordinate axes, which is
analogous to the procedure we described.  During the sampling steps, VEGAS uses
the information it obtained to continuously improve the weight function.

To have a meaningful comparison, we will add physics information in the form of
the $\vec{u} \to \vec{v}$ coordinate mapping (see above). The form
\eqnref{eq:warper1d_app}  is only a good approximation to our integrand in the
$\vec{v}$ variables. Although in our case VEGAS converges poorly in both
$\vec{u}$ and $\vec{v}$ variables, it is far worse in $\vec{u}$.

Our approach still differs from VEGAS in two essential ways. First, a key
feature of our approach is to correctly capture the decaying asymptotic
structure in the model function analytically.  As we have shown in
Fig.~\ref{fig:f-tau}, a pure projection method is prohibitively inefficient in
sampling the long tails and leads to poor convergence. It is only useful as a
small correction to a good starting point. Second, the continuous change of
model function of VEGAS cannot be used together with low-discrepancy sequences,
since it breaks their special properties. For QQMC to work and achieve $1/N$
convergence, it is essential to construct the model function entirely before the
full calculation is performed. Samples cannot be ``reused'' as in Monte Carlo
sampling. This again requires an numerically efficient way to construct the
model function which is not a priori provided by VEGAS.

Finally, we note that the idea of the model function is more general than the
product form of  \eqnref{eq:warper1d_app} and VEGAS. Although they were
sufficient for the Anderson impurity model calculations presented here, more
general model function such as an MPS \eqnref{eq:warperMPS} are important for
more complex many-body systems.

\bibliography{warper-autobib,warper-manual}

\begin{thebibliography}{60}%
\makeatletter
\providecommand \@ifxundefined [1]{%
 \@ifx{#1\undefined}
}%
\providecommand \@ifnum [1]{%
 \ifnum #1\expandafter \@firstoftwo
 \else \expandafter \@secondoftwo
 \fi
}%
\providecommand \@ifx [1]{%
 \ifx #1\expandafter \@firstoftwo
 \else \expandafter \@secondoftwo
 \fi
}%
\providecommand \natexlab [1]{#1}%
\providecommand \enquote  [1]{``#1''}%
\providecommand \bibnamefont  [1]{#1}%
\providecommand \bibfnamefont [1]{#1}%
\providecommand \citenamefont [1]{#1}%
\providecommand \href@noop [0]{\@secondoftwo}%
\providecommand \href [0]{\begingroup \@sanitize@url \@href}%
\providecommand \@href[1]{\@@startlink{#1}\@@href}%
\providecommand \@@href[1]{\endgroup#1\@@endlink}%
\providecommand \@sanitize@url [0]{\catcode `\\12\catcode `\$12\catcode
  `\&12\catcode `\#12\catcode `\^12\catcode `\_12\catcode `\%12\relax}%
\providecommand \@@startlink[1]{}%
\providecommand \@@endlink[0]{}%
\providecommand \url  [0]{\begingroup\@sanitize@url \@url }%
\providecommand \@url [1]{\endgroup\@href {#1}{\urlprefix }}%
\providecommand \urlprefix  [0]{URL }%
\providecommand \Eprint [0]{\href }%
\providecommand \doibase [0]{http://dx.doi.org/}%
\providecommand \selectlanguage [0]{\@gobble}%
\providecommand \bibinfo  [0]{\@secondoftwo}%
\providecommand \bibfield  [0]{\@secondoftwo}%
\providecommand \translation [1]{[#1]}%
\providecommand \BibitemOpen [0]{}%
\providecommand \bibitemStop [0]{}%
\providecommand \bibitemNoStop [0]{.\EOS\space}%
\providecommand \EOS [0]{\spacefactor3000\relax}%
\providecommand \BibitemShut  [1]{\csname bibitem#1\endcsname}%
\let\auto@bib@innerbib\@empty
\bibitem [{\citenamefont {Gross}\ and\ \citenamefont
  {Bloch}(2017)}]{Gross_Bloch_2017}%
  \BibitemOpen
  \bibfield  {author} {\bibinfo {author} {\bibfnamefont {Christian}\
  \bibnamefont {Gross}}\ and\ \bibinfo {author} {\bibfnamefont {Immanuel}\
  \bibnamefont {Bloch}},\ }\bibfield  {title} {\enquote {\bibinfo {title}
  {Quantum simulations with ultracold atoms in optical lattices},}\ }\href
  {\doibase 10.1126/science.aal3837} {\bibfield  {journal} {\bibinfo  {journal}
  {Science}\ }\textbf {\bibinfo {volume} {357}},\ \bibinfo {pages} {995}
  (\bibinfo {year} {2017})}\BibitemShut {NoStop}%
\bibitem [{\citenamefont {Blatt}\ and\ \citenamefont
  {Roos}(2012)}]{Blatt_Ross_TrappedIons_2012}%
  \BibitemOpen
  \bibfield  {author} {\bibinfo {author} {\bibfnamefont {R.}~\bibnamefont
  {Blatt}}\ and\ \bibinfo {author} {\bibfnamefont {C.~F.}\ \bibnamefont
  {Roos}},\ }\bibfield  {title} {\enquote {\bibinfo {title} {Quantum
  simulations with trapped ions},}\ }\href {\doibase 10.1038/NPHYS2252}
  {\bibfield  {journal} {\bibinfo  {journal} {Nature Phys}\ }\textbf {\bibinfo
  {volume} {8}},\ \bibinfo {pages} {277} (\bibinfo {year} {2012})}\BibitemShut
  {NoStop}%
\bibitem [{\citenamefont {Goldhaber-Gordon}\ \emph
  {et~al.}(1998{\natexlab{a}})\citenamefont {Goldhaber-Gordon}, \citenamefont
  {Shtrikman}, \citenamefont {Mahalu}, \citenamefont {Abusch-Magder},
  \citenamefont {Meirav},\ and\ \citenamefont
  {Kastner}}]{Goldhaber-Gordon1998}%
  \BibitemOpen
  \bibfield  {author} {\bibinfo {author} {\bibfnamefont {D.}~\bibnamefont
  {Goldhaber-Gordon}}, \bibinfo {author} {\bibfnamefont {Hadas}\ \bibnamefont
  {Shtrikman}}, \bibinfo {author} {\bibfnamefont {D.}~\bibnamefont {Mahalu}},
  \bibinfo {author} {\bibfnamefont {David}\ \bibnamefont {Abusch-Magder}},
  \bibinfo {author} {\bibfnamefont {U.}~\bibnamefont {Meirav}}, \ and\ \bibinfo
  {author} {\bibfnamefont {M.~A.}\ \bibnamefont {Kastner}},\ }\bibfield
  {title} {\enquote {\bibinfo {title} {{Kondo} effect in a single-electron
  transistor},}\ }\href {https://doi.org/10.1038/34373} {\bibfield  {journal}
  {\bibinfo  {journal} {Nature}\ }\textbf {\bibinfo {volume} {391}},\ \bibinfo
  {pages} {156} (\bibinfo {year} {1998}{\natexlab{a}})}\BibitemShut {NoStop}%
\bibitem [{\citenamefont {Goldhaber-Gordon}\ \emph
  {et~al.}(1998{\natexlab{b}})\citenamefont {Goldhaber-Gordon}, \citenamefont
  {G\"ores}, \citenamefont {Kastner}, \citenamefont {Shtrikman}, \citenamefont
  {Mahalu},\ and\ \citenamefont {Meirav}}]{Goldhaber-Gordon1998a}%
  \BibitemOpen
  \bibfield  {author} {\bibinfo {author} {\bibfnamefont {D.}~\bibnamefont
  {Goldhaber-Gordon}}, \bibinfo {author} {\bibfnamefont {J.}~\bibnamefont
  {G\"ores}}, \bibinfo {author} {\bibfnamefont {M.~A.}\ \bibnamefont
  {Kastner}}, \bibinfo {author} {\bibfnamefont {Hadas}\ \bibnamefont
  {Shtrikman}}, \bibinfo {author} {\bibfnamefont {D.}~\bibnamefont {Mahalu}}, \
  and\ \bibinfo {author} {\bibfnamefont {U.}~\bibnamefont {Meirav}},\
  }\bibfield  {title} {\enquote {\bibinfo {title} {From the {Kondo} regime to
  the mixed-valence regime in a single-electron transistor},}\ }\href {\doibase
  10.1103/PhysRevLett.81.5225} {\bibfield  {journal} {\bibinfo  {journal}
  {Phys. Rev. Lett.}\ }\textbf {\bibinfo {volume} {81}},\ \bibinfo {pages}
  {5225--5228} (\bibinfo {year} {1998}{\natexlab{b}})}\BibitemShut {NoStop}%
\bibitem [{\citenamefont {Cronenwett}\ \emph {et~al.}(1998)\citenamefont
  {Cronenwett}, \citenamefont {Oosterkamp},\ and\ \citenamefont
  {Kouwenhoven}}]{Kouwenhoven_Kondo_2000}%
  \BibitemOpen
  \bibfield  {author} {\bibinfo {author} {\bibfnamefont {Sara~M.}\ \bibnamefont
  {Cronenwett}}, \bibinfo {author} {\bibfnamefont {Tjerk~H.}\ \bibnamefont
  {Oosterkamp}}, \ and\ \bibinfo {author} {\bibfnamefont {Leo~P.}\ \bibnamefont
  {Kouwenhoven}},\ }\bibfield  {title} {\enquote {\bibinfo {title} {A tunable
  {Kondo} effect in quantum dots},}\ }\href {\doibase
  10.1126/science.281.5376.540} {\bibfield  {journal} {\bibinfo  {journal}
  {Science}\ }\textbf {\bibinfo {volume} {281}},\ \bibinfo {pages} {540--544}
  (\bibinfo {year} {1998})}\BibitemShut {NoStop}%
\bibitem [{\citenamefont {Iftikhar}\ \emph {et~al.}(2018)\citenamefont
  {Iftikhar}, \citenamefont {Anthore}, \citenamefont {Mitchell}, \citenamefont
  {Parmentier}, \citenamefont {Gennser}, \citenamefont {Ouerghi}, \citenamefont
  {Cavanna}, \citenamefont {Mora}, \citenamefont {Simon},\ and\ \citenamefont
  {Pierre}}]{Pierre_Science_Kondo_2018}%
  \BibitemOpen
  \bibfield  {author} {\bibinfo {author} {\bibfnamefont {Z.}~\bibnamefont
  {Iftikhar}}, \bibinfo {author} {\bibfnamefont {A.}~\bibnamefont {Anthore}},
  \bibinfo {author} {\bibfnamefont {A.~K.}\ \bibnamefont {Mitchell}}, \bibinfo
  {author} {\bibfnamefont {F.~D.}\ \bibnamefont {Parmentier}}, \bibinfo
  {author} {\bibfnamefont {U.}~\bibnamefont {Gennser}}, \bibinfo {author}
  {\bibfnamefont {A.}~\bibnamefont {Ouerghi}}, \bibinfo {author} {\bibfnamefont
  {A.}~\bibnamefont {Cavanna}}, \bibinfo {author} {\bibfnamefont
  {C.}~\bibnamefont {Mora}}, \bibinfo {author} {\bibfnamefont {P.}~\bibnamefont
  {Simon}}, \ and\ \bibinfo {author} {\bibfnamefont {F.}~\bibnamefont
  {Pierre}},\ }\bibfield  {title} {\enquote {\bibinfo {title} {Tunable quantum
  criticality and super-ballistic transport in a “charge” {Kondo}
  circuit},}\ }\href {\doibase 10.1126/science.aan5592} {\bibfield  {journal}
  {\bibinfo  {journal} {Science}\ }\textbf {\bibinfo {volume} {360}},\ \bibinfo
  {pages} {1315} (\bibinfo {year} {2018})}\BibitemShut {NoStop}%
\bibitem [{\citenamefont {Bernien}\ \emph {et~al.}(2017)\citenamefont
  {Bernien}, \citenamefont {Schwartz}, \citenamefont {Keesling}, \citenamefont
  {Levine}, \citenamefont {Omran}, \citenamefont {Pichler}, \citenamefont
  {Choi}, \citenamefont {Zibrov}, \citenamefont {Endres}, \citenamefont
  {Greiner}, \citenamefont {Vuletić},\ and\ \citenamefont
  {Lukin}}]{Lukin_51_atoms_2017}%
  \BibitemOpen
  \bibfield  {author} {\bibinfo {author} {\bibfnamefont {Hannes}\ \bibnamefont
  {Bernien}}, \bibinfo {author} {\bibfnamefont {Sylvain}\ \bibnamefont
  {Schwartz}}, \bibinfo {author} {\bibfnamefont {Alexander}\ \bibnamefont
  {Keesling}}, \bibinfo {author} {\bibfnamefont {Harry}\ \bibnamefont
  {Levine}}, \bibinfo {author} {\bibfnamefont {Ahmed}\ \bibnamefont {Omran}},
  \bibinfo {author} {\bibfnamefont {Hannes}\ \bibnamefont {Pichler}}, \bibinfo
  {author} {\bibfnamefont {Soonwon}\ \bibnamefont {Choi}}, \bibinfo {author}
  {\bibfnamefont {Alexander~S.}\ \bibnamefont {Zibrov}}, \bibinfo {author}
  {\bibfnamefont {Manuel}\ \bibnamefont {Endres}}, \bibinfo {author}
  {\bibfnamefont {Markus}\ \bibnamefont {Greiner}}, \bibinfo {author}
  {\bibfnamefont {Vladan}\ \bibnamefont {Vuletić}}, \ and\ \bibinfo {author}
  {\bibfnamefont {Mikhail~D.}\ \bibnamefont {Lukin}},\ }\bibfield  {title}
  {\enquote {\bibinfo {title} {Probing many-body dynamics on a 51-atom quantum
  simulator},}\ }\href {\doibase 10.1038/nature24622} {\bibfield  {journal}
  {\bibinfo  {journal} {Nature}\ }\textbf {\bibinfo {volume} {551}},\ \bibinfo
  {pages} {579} (\bibinfo {year} {2017})}\BibitemShut {NoStop}%
\bibitem [{\citenamefont {LeBlanc}\ \emph {et~al.}(2015)\citenamefont
  {LeBlanc}, \citenamefont {Antipov}, \citenamefont {Becca}, \citenamefont
  {Bulik}, \citenamefont {Chan}, \citenamefont {Chung}, \citenamefont {Deng},
  \citenamefont {Ferrero}, \citenamefont {Henderson}, \citenamefont
  {Jim\'enez-Hoyos}, \citenamefont {Kozik}, \citenamefont {Liu}, \citenamefont
  {Millis}, \citenamefont {Prokof'ev}, \citenamefont {Qin}, \citenamefont
  {Scuseria}, \citenamefont {Shi}, \citenamefont {Svistunov}, \citenamefont
  {Tocchio}, \citenamefont {Tupitsyn}, \citenamefont {White}, \citenamefont
  {Zhang}, \citenamefont {Zheng}, \citenamefont {Zhu},\ and\ \citenamefont
  {Gull}}]{LeblancPRXSimonsColl_2015}%
  \BibitemOpen
  \bibfield  {author} {\bibinfo {author} {\bibfnamefont {J.~P.~F.}\
  \bibnamefont {LeBlanc}}, \bibinfo {author} {\bibfnamefont {Andrey~E.}\
  \bibnamefont {Antipov}}, \bibinfo {author} {\bibfnamefont {Federico}\
  \bibnamefont {Becca}}, \bibinfo {author} {\bibfnamefont {Ireneusz~W.}\
  \bibnamefont {Bulik}}, \bibinfo {author} {\bibfnamefont {Garnet Kin-Lic}\
  \bibnamefont {Chan}}, \bibinfo {author} {\bibfnamefont {Chia-Min}\
  \bibnamefont {Chung}}, \bibinfo {author} {\bibfnamefont {Youjin}\
  \bibnamefont {Deng}}, \bibinfo {author} {\bibfnamefont {Michel}\ \bibnamefont
  {Ferrero}}, \bibinfo {author} {\bibfnamefont {Thomas~M.}\ \bibnamefont
  {Henderson}}, \bibinfo {author} {\bibfnamefont {Carlos~A.}\ \bibnamefont
  {Jim\'enez-Hoyos}}, \bibinfo {author} {\bibfnamefont {E.}~\bibnamefont
  {Kozik}}, \bibinfo {author} {\bibfnamefont {Xuan-Wen}\ \bibnamefont {Liu}},
  \bibinfo {author} {\bibfnamefont {Andrew~J.}\ \bibnamefont {Millis}},
  \bibinfo {author} {\bibfnamefont {N.~V.}\ \bibnamefont {Prokof'ev}}, \bibinfo
  {author} {\bibfnamefont {Mingpu}\ \bibnamefont {Qin}}, \bibinfo {author}
  {\bibfnamefont {Gustavo~E.}\ \bibnamefont {Scuseria}}, \bibinfo {author}
  {\bibfnamefont {Hao}\ \bibnamefont {Shi}}, \bibinfo {author} {\bibfnamefont
  {B.~V.}\ \bibnamefont {Svistunov}}, \bibinfo {author} {\bibfnamefont
  {Luca~F.}\ \bibnamefont {Tocchio}}, \bibinfo {author} {\bibfnamefont {I.~S.}\
  \bibnamefont {Tupitsyn}}, \bibinfo {author} {\bibfnamefont {Steven~R.}\
  \bibnamefont {White}}, \bibinfo {author} {\bibfnamefont {Shiwei}\
  \bibnamefont {Zhang}}, \bibinfo {author} {\bibfnamefont {Bo-Xiao}\
  \bibnamefont {Zheng}}, \bibinfo {author} {\bibfnamefont {Zhenyue}\
  \bibnamefont {Zhu}}, \ and\ \bibinfo {author} {\bibfnamefont {Emanuel}\
  \bibnamefont {Gull}} (\bibinfo {collaboration} {Simons Collaboration on the
  Many-Electron Problem}),\ }\bibfield  {title} {\enquote {\bibinfo {title}
  {Solutions of the two-dimensional {Hubbard} model: Benchmarks and results
  from a wide range of numerical algorithms},}\ }\href {\doibase
  10.1103/PhysRevX.5.041041} {\bibfield  {journal} {\bibinfo  {journal} {Phys.
  Rev. X}\ }\textbf {\bibinfo {volume} {5}},\ \bibinfo {pages} {041041}
  (\bibinfo {year} {2015})}\BibitemShut {NoStop}%
\bibitem [{\citenamefont {Cohen}\ \emph {et~al.}(2015)\citenamefont {Cohen},
  \citenamefont {Gull}, \citenamefont {Reichman},\ and\ \citenamefont
  {Millis}}]{Cohen2015}%
  \BibitemOpen
  \bibfield  {author} {\bibinfo {author} {\bibfnamefont {Guy}\ \bibnamefont
  {Cohen}}, \bibinfo {author} {\bibfnamefont {Emanuel}\ \bibnamefont {Gull}},
  \bibinfo {author} {\bibfnamefont {David~R.}\ \bibnamefont {Reichman}}, \ and\
  \bibinfo {author} {\bibfnamefont {Andrew~J.}\ \bibnamefont {Millis}},\
  }\bibfield  {title} {\enquote {\bibinfo {title} {Taming the dynamical sign
  problem in real-time evolution of quantum many-body problems},}\ }\href
  {\doibase 10.1103/PhysRevLett.115.266802} {\bibfield  {journal} {\bibinfo
  {journal} {Phys. Rev. Lett.}\ }\textbf {\bibinfo {volume} {115}},\ \bibinfo
  {pages} {266802} (\bibinfo {year} {2015})}\BibitemShut {NoStop}%
\bibitem [{\citenamefont {Profumo}\ \emph {et~al.}(2015)\citenamefont
  {Profumo}, \citenamefont {Groth}, \citenamefont {Messio}, \citenamefont
  {Parcollet},\ and\ \citenamefont {Waintal}}]{Profumo_1504}%
  \BibitemOpen
  \bibfield  {author} {\bibinfo {author} {\bibfnamefont {Rosario E.~V.}\
  \bibnamefont {Profumo}}, \bibinfo {author} {\bibfnamefont {Christoph}\
  \bibnamefont {Groth}}, \bibinfo {author} {\bibfnamefont {Laura}\ \bibnamefont
  {Messio}}, \bibinfo {author} {\bibfnamefont {Olivier}\ \bibnamefont
  {Parcollet}}, \ and\ \bibinfo {author} {\bibfnamefont {Xavier}\ \bibnamefont
  {Waintal}},\ }\bibfield  {title} {\enquote {\bibinfo {title} {Quantum {Monte}
  {Carlo} for correlated out-of-equilibrium nanoelectronic devices},}\ }\href
  {\doibase 10.1103/PhysRevB.91.245154} {\bibfield  {journal} {\bibinfo
  {journal} {Phys. Rev. B}\ }\textbf {\bibinfo {volume} {91}},\ \bibinfo
  {pages} {245154} (\bibinfo {year} {2015})},\ \Eprint
  {http://arxiv.org/abs/1504.02132} {arXiv:1504.02132} \BibitemShut {NoStop}%
\bibitem [{\citenamefont {Bertrand}\ \emph
  {et~al.}(2019{\natexlab{a}})\citenamefont {Bertrand}, \citenamefont
  {Florens}, \citenamefont {Parcollet},\ and\ \citenamefont
  {Waintal}}]{Bertrand_1903_series}%
  \BibitemOpen
  \bibfield  {author} {\bibinfo {author} {\bibfnamefont {Corentin}\
  \bibnamefont {Bertrand}}, \bibinfo {author} {\bibfnamefont {Serge}\
  \bibnamefont {Florens}}, \bibinfo {author} {\bibfnamefont {Olivier}\
  \bibnamefont {Parcollet}}, \ and\ \bibinfo {author} {\bibfnamefont {Xavier}\
  \bibnamefont {Waintal}},\ }\bibfield  {title} {\enquote {\bibinfo {title}
  {Reconstructing nonequilibrium regimes of quantum many-body systems from the
  analytical structure of perturbative expansions},}\ }\href {\doibase
  10.1103/PhysRevX.9.041008} {\bibfield  {journal} {\bibinfo  {journal} {Phys.
  Rev. X}\ }\textbf {\bibinfo {volume} {9}},\ \bibinfo {pages} {041008}
  (\bibinfo {year} {2019}{\natexlab{a}})},\ \Eprint
  {http://arxiv.org/abs/1903.11646} {arXiv:1903.11646} \BibitemShut {NoStop}%
\bibitem [{\citenamefont {Prokof'ev}\ and\ \citenamefont
  {Svistunov}(1998)}]{Prokofev_9804}%
  \BibitemOpen
  \bibfield  {author} {\bibinfo {author} {\bibfnamefont {Nikolai~V.}\
  \bibnamefont {Prokof'ev}}\ and\ \bibinfo {author} {\bibfnamefont {Boris~V.}\
  \bibnamefont {Svistunov}},\ }\bibfield  {title} {\enquote {\bibinfo {title}
  {Polaron problem by diagrammatic quantum {Monte} {Carlo}},}\ }\href {\doibase
  10.1103/PhysRevLett.81.2514} {\bibfield  {journal} {\bibinfo  {journal}
  {Phys. Rev. Lett.}\ }\textbf {\bibinfo {volume} {81}},\ \bibinfo {pages}
  {2514} (\bibinfo {year} {1998})},\ \Eprint
  {http://arxiv.org/abs/cond-mat/9804097} {arXiv:cond-mat/9804097} \BibitemShut
  {NoStop}%
\bibitem [{\citenamefont {Prokof’ev}\ and\ \citenamefont
  {Svistunov}(2008)}]{Prokofev_0801}%
  \BibitemOpen
  \bibfield  {author} {\bibinfo {author} {\bibfnamefont {N.~V.}\ \bibnamefont
  {Prokof’ev}}\ and\ \bibinfo {author} {\bibfnamefont {B.~V.}\ \bibnamefont
  {Svistunov}},\ }\bibfield  {title} {\enquote {\bibinfo {title} {Bold
  diagrammatic {Monte} {Carlo}: A generic sign-problem tolerant technique for
  polaron models and possibly interacting many-body problems},}\ }\href
  {\doibase 10.1103/PhysRevB.77.125101} {\bibfield  {journal} {\bibinfo
  {journal} {Phys. Rev. B}\ }\textbf {\bibinfo {volume} {77}},\ \bibinfo
  {pages} {125101} (\bibinfo {year} {2008})},\ \Eprint
  {http://arxiv.org/abs/0801.0911} {arXiv:0801.0911} \BibitemShut {NoStop}%
\bibitem [{\citenamefont {Mishchenko}\ \emph {et~al.}(2001)\citenamefont
  {Mishchenko}, \citenamefont {Prokof'ev}, \citenamefont {Svistunov},\ and\
  \citenamefont {Sakamoto}}]{Mishchenko_9910}%
  \BibitemOpen
  \bibfield  {author} {\bibinfo {author} {\bibfnamefont {A.~S.}\ \bibnamefont
  {Mishchenko}}, \bibinfo {author} {\bibfnamefont {N.~V.}\ \bibnamefont
  {Prokof'ev}}, \bibinfo {author} {\bibfnamefont {B.~V.}\ \bibnamefont
  {Svistunov}}, \ and\ \bibinfo {author} {\bibfnamefont {A.}~\bibnamefont
  {Sakamoto}},\ }\bibfield  {title} {\enquote {\bibinfo {title} {Comprehensive
  study of {Fröhlich} polaron},}\ }\href {\doibase 10.1142/S0217979201009050}
  {\bibfield  {journal} {\bibinfo  {journal} {Int. J. Mod. Phys. B}\ }\textbf
  {\bibinfo {volume} {15}},\ \bibinfo {pages} {3940--3943} (\bibinfo {year}
  {2001})}\BibitemShut {NoStop}%
\bibitem [{\citenamefont {Van~Houcke}\ \emph {et~al.}(2012)\citenamefont
  {Van~Houcke}, \citenamefont {Werner}, \citenamefont {Kozik}, \citenamefont
  {Prokof’ev}, \citenamefont {Svistunov}, \citenamefont {Ku}, \citenamefont
  {Sommer}, \citenamefont {Cheuk}, \citenamefont {Schirotzek},\ and\
  \citenamefont {Zwierlein}}]{VanHoucke_1110}%
  \BibitemOpen
  \bibfield  {author} {\bibinfo {author} {\bibfnamefont {K.}~\bibnamefont
  {Van~Houcke}}, \bibinfo {author} {\bibfnamefont {F.}~\bibnamefont {Werner}},
  \bibinfo {author} {\bibfnamefont {E.}~\bibnamefont {Kozik}}, \bibinfo
  {author} {\bibfnamefont {N.}~\bibnamefont {Prokof’ev}}, \bibinfo {author}
  {\bibfnamefont {B.}~\bibnamefont {Svistunov}}, \bibinfo {author}
  {\bibfnamefont {M.~J.~H.}\ \bibnamefont {Ku}}, \bibinfo {author}
  {\bibfnamefont {A.~T.}\ \bibnamefont {Sommer}}, \bibinfo {author}
  {\bibfnamefont {L.~W.}\ \bibnamefont {Cheuk}}, \bibinfo {author}
  {\bibfnamefont {A.}~\bibnamefont {Schirotzek}}, \ and\ \bibinfo {author}
  {\bibfnamefont {M.~W.}\ \bibnamefont {Zwierlein}},\ }\bibfield  {title}
  {\enquote {\bibinfo {title} {Feynman diagrams versus {Fermi}-gas
  {Feynman} emulator},}\ }\href {\doibase 10.1038/NPHYS2273} {\bibfield
  {journal} {\bibinfo  {journal} {Nature Phys}\ }\textbf {\bibinfo {volume}
  {8}},\ \bibinfo {pages} {366} (\bibinfo {year} {2012})},\ \Eprint
  {http://arxiv.org/abs/1110.3747} {arXiv:1110.3747} \BibitemShut {NoStop}%
\bibitem [{\citenamefont {Wu}\ \emph {et~al.}(2017)\citenamefont {Wu},
  \citenamefont {Ferrero}, \citenamefont {Georges},\ and\ \citenamefont
  {Kozik}}]{Wu_1608}%
  \BibitemOpen
  \bibfield  {author} {\bibinfo {author} {\bibfnamefont {Wei}\ \bibnamefont
  {Wu}}, \bibinfo {author} {\bibfnamefont {Michel}\ \bibnamefont {Ferrero}},
  \bibinfo {author} {\bibfnamefont {Antoine}\ \bibnamefont {Georges}}, \ and\
  \bibinfo {author} {\bibfnamefont {Evgeny}\ \bibnamefont {Kozik}},\ }\bibfield
   {title} {\enquote {\bibinfo {title} {Controlling {Feynman} diagrammatic
  expansions: Physical nature of the pseudogap in the two-dimensional {Hubbard}
  model},}\ }\href {\doibase 10.1103/PhysRevB.96.041105} {\bibfield  {journal}
  {\bibinfo  {journal} {Phys. Rev. B}\ }\textbf {\bibinfo {volume} {96}},\
  \bibinfo {pages} {041105} (\bibinfo {year} {2017})},\ \Eprint
  {http://arxiv.org/abs/1608.08402} {arXiv:1608.08402} \BibitemShut {NoStop}%
\bibitem [{\citenamefont {Rossi}(2017)}]{Rossi_1612}%
  \BibitemOpen
  \bibfield  {author} {\bibinfo {author} {\bibfnamefont {Riccardo}\
  \bibnamefont {Rossi}},\ }\bibfield  {title} {\enquote {\bibinfo {title}
  {Determinant diagrammatic {Monte} {Carlo} algorithm in the thermodynamic
  limit},}\ }\href {\doibase 10.1103/PhysRevLett.119.045701} {\bibfield
  {journal} {\bibinfo  {journal} {Phys. Rev. Lett.}\ }\textbf {\bibinfo
  {volume} {119}},\ \bibinfo {pages} {045701} (\bibinfo {year} {2017})},\
  \Eprint {http://arxiv.org/abs/1612.05184} {arXiv:1612.05184} \BibitemShut
  {NoStop}%
\bibitem [{\citenamefont {Chen}\ and\ \citenamefont {Haule}(2019)}]{Chen_1809}%
  \BibitemOpen
  \bibfield  {author} {\bibinfo {author} {\bibfnamefont {Kun}\ \bibnamefont
  {Chen}}\ and\ \bibinfo {author} {\bibfnamefont {Kristjan}\ \bibnamefont
  {Haule}},\ }\bibfield  {title} {\enquote {\bibinfo {title} {A combined
  variational and diagrammatic quantum {Monte} {Carlo} approach to the
  many-electron problem},}\ }\href {\doibase 10.1038/s41467-019-11708-6}
  {\bibfield  {journal} {\bibinfo  {journal} {Nat Commun}\ }\textbf {\bibinfo
  {volume} {10}},\ \bibinfo {pages} {3725} (\bibinfo {year} {2019})},\ \Eprint
  {http://arxiv.org/abs/1809.04651} {arXiv:1809.04651} \BibitemShut {NoStop}%
\bibitem [{\citenamefont {Bertrand}\ \emph
  {et~al.}(2019{\natexlab{b}})\citenamefont {Bertrand}, \citenamefont
  {Parcollet}, \citenamefont {Maillard},\ and\ \citenamefont
  {Waintal}}]{Bertrand_1903_kernel}%
  \BibitemOpen
  \bibfield  {author} {\bibinfo {author} {\bibfnamefont {Corentin}\
  \bibnamefont {Bertrand}}, \bibinfo {author} {\bibfnamefont {Olivier}\
  \bibnamefont {Parcollet}}, \bibinfo {author} {\bibfnamefont {Antoine}\
  \bibnamefont {Maillard}}, \ and\ \bibinfo {author} {\bibfnamefont {Xavier}\
  \bibnamefont {Waintal}},\ }\bibfield  {title} {\enquote {\bibinfo {title}
  {Quantum {Monte} {Carlo} algorithm for out-of-equilibrium {Green}'s functions
  at long times},}\ }\href {\doibase 10.1103/PhysRevB.100.125129} {\bibfield
  {journal} {\bibinfo  {journal} {Phys. Rev. B}\ }\textbf {\bibinfo {volume}
  {100}},\ \bibinfo {pages} {125129} (\bibinfo {year} {2019}{\natexlab{b}})},\
  \Eprint {http://arxiv.org/abs/1903.11636} {arXiv:1903.11636} \BibitemShut
  {NoStop}%
\bibitem [{\citenamefont {Moutenet}\ \emph {et~al.}(2019)\citenamefont
  {Moutenet}, \citenamefont {Seth}, \citenamefont {Ferrero},\ and\
  \citenamefont {Parcollet}}]{Moutenet_1904}%
  \BibitemOpen
  \bibfield  {author} {\bibinfo {author} {\bibfnamefont {Alice}\ \bibnamefont
  {Moutenet}}, \bibinfo {author} {\bibfnamefont {Priyanka}\ \bibnamefont
  {Seth}}, \bibinfo {author} {\bibfnamefont {Michel}\ \bibnamefont {Ferrero}},
  \ and\ \bibinfo {author} {\bibfnamefont {Olivier}\ \bibnamefont
  {Parcollet}},\ }\bibfield  {title} {\enquote {\bibinfo {title} {Cancellation
  of vacuum diagrams and the long-time limit in out-of-equilibrium diagrammatic
  quantum {Monte} {Carlo}},}\ }\href {\doibase 10.1103/PhysRevB.100.085125}
  {\bibfield  {journal} {\bibinfo  {journal} {Phys. Rev. B}\ }\textbf {\bibinfo
  {volume} {100}},\ \bibinfo {pages} {085125} (\bibinfo {year} {2019})},\
  \Eprint {http://arxiv.org/abs/1904.11969} {arXiv:1904.11969} \BibitemShut
  {NoStop}%
\bibitem [{\citenamefont {Rossi}\ \emph {et~al.}()\citenamefont {Rossi},
  \citenamefont {Simkovic},\ and\ \citenamefont {Ferrero}}]{Rossi_2001}%
  \BibitemOpen
  \bibfield  {author} {\bibinfo {author} {\bibfnamefont {Riccardo}\
  \bibnamefont {Rossi}}, \bibinfo {author} {\bibfnamefont {Fedor}\ \bibnamefont
  {Simkovic}}, \ and\ \bibinfo {author} {\bibfnamefont {Michel}\ \bibnamefont
  {Ferrero}},\ }\bibfield  {title} {\enquote {\bibinfo {title} {Renormalized
  perturbation theory at large expansion orders},}\ }\href@noop {} {\ }\Eprint
  {http://arxiv.org/abs/2001.09133} {arXiv:2001.09133} \BibitemShut {NoStop}%
\bibitem [{\citenamefont {Dick}\ \emph {et~al.}(2013)\citenamefont {Dick},
  \citenamefont {Kuo},\ and\ \citenamefont {Sloan}}]{Dick_2013}%
  \BibitemOpen
  \bibfield  {author} {\bibinfo {author} {\bibfnamefont {Josef}\ \bibnamefont
  {Dick}}, \bibinfo {author} {\bibfnamefont {Frances~Y.}\ \bibnamefont {Kuo}},
  \ and\ \bibinfo {author} {\bibfnamefont {Ian~H.}\ \bibnamefont {Sloan}},\
  }\bibfield  {title} {\enquote {\bibinfo {title} {High-dimensional
  integration: The quasi-{Monte} {Carlo} way},}\ }\href {\doibase
  10.1017/S0962492913000044} {\bibfield  {journal} {\bibinfo  {journal} {Acta
  Numer.}\ }\textbf {\bibinfo {volume} {22}},\ \bibinfo {pages} {133} (\bibinfo
  {year} {2013})}\BibitemShut {NoStop}%
\bibitem [{\citenamefont {Nuyens}()}]{Nuyens_1308}%
  \BibitemOpen
  \bibfield  {author} {\bibinfo {author} {\bibfnamefont {Dirk}\ \bibnamefont
  {Nuyens}},\ }\bibfield  {title} {\enquote {\bibinfo {title} {The construction
  of good lattice rules and polynomial lattice rules},}\ }\href@noop {} {\
  }\Eprint {http://arxiv.org/abs/1308.3601} {arXiv:1308.3601} \BibitemShut
  {NoStop}%
\bibitem [{\citenamefont {Dick}\ and\ \citenamefont
  {Pillichshammer}(2010)}]{Dick_Pillichshammer_2010}%
  \BibitemOpen
  \bibfield  {author} {\bibinfo {author} {\bibfnamefont {Josef}\ \bibnamefont
  {Dick}}\ and\ \bibinfo {author} {\bibfnamefont {Friedrich}\ \bibnamefont
  {Pillichshammer}},\ }\href {\doibase 10.1017/CBO9780511761188} {\emph
  {\bibinfo {title} {Digital Nets and Sequences: Discrepancy Theory and
  Quasi–{Monte} {Carlo} Integration}}}\ (\bibinfo  {publisher} {Cambridge
  University Press},\ \bibinfo {address} {Cambridge},\ \bibinfo {year}
  {2010})\BibitemShut {NoStop}%
\bibitem [{\citenamefont {L'Ecuyer}(2018)}]{Lecuyer_2018}%
  \BibitemOpen
  \bibfield  {author} {\bibinfo {author} {\bibfnamefont {Pierre}\ \bibnamefont
  {L'Ecuyer}},\ }\bibfield  {title} {\enquote {\bibinfo {title} {Randomized
  quasi-{Monte} {Carlo}: An introduction for practitioners},}\ }in\ \href
  {\doibase https://doi.org/10.1007/978-3-319-91436-7} {\emph {\bibinfo
  {booktitle} {Monte Carlo and Quasi-Monte Carlo Methods}}},\ \bibinfo {editor}
  {edited by\ \bibinfo {editor} {\bibfnamefont {Art~B.}\ \bibnamefont {Owen}}\
  and\ \bibinfo {editor} {\bibfnamefont {Peter~W.}\ \bibnamefont {Glynn}}}\
  (\bibinfo  {publisher} {Springer International Publishing},\ \bibinfo
  {address} {Cham},\ \bibinfo {year} {2018})\ pp.\ \bibinfo {pages}
  {29--52}\BibitemShut {NoStop}%
\bibitem [{\citenamefont {Sobol'}(1967)}]{Sobol_1967}%
  \BibitemOpen
  \bibfield  {author} {\bibinfo {author} {\bibfnamefont {I.M}\ \bibnamefont
  {Sobol'}},\ }\bibfield  {title} {\enquote {\bibinfo {title} {On the
  distribution of points in a cube and the approximate evaluation of
  integrals},}\ }\href {\doibase 10.1016/0041-5553(67)90144-9} {\bibfield
  {journal} {\bibinfo  {journal} {USSR Computational Mathematics and
  Mathematical Physics}\ }\textbf {\bibinfo {volume} {7}},\ \bibinfo {pages}
  {86} (\bibinfo {year} {1967})}\BibitemShut {NoStop}%
\bibitem [{\citenamefont {Kuo}\ and\ \citenamefont {Nuyens}(2016)}]{Kuo_1606}%
  \BibitemOpen
  \bibfield  {author} {\bibinfo {author} {\bibfnamefont {Frances~Y.}\
  \bibnamefont {Kuo}}\ and\ \bibinfo {author} {\bibfnamefont {Dirk}\
  \bibnamefont {Nuyens}},\ }\bibfield  {title} {\enquote {\bibinfo {title}
  {Application of quasi-{Monte} {Carlo} methods to elliptic pdes with random
  diffusion coefficients: A survey of analysis and implementation},}\ }\href
  {\doibase 10.1007/s10208-016-9329-5} {\bibfield  {journal} {\bibinfo
  {journal} {Found Comput Math}\ }\textbf {\bibinfo {volume} {16}},\ \bibinfo
  {pages} {1631} (\bibinfo {year} {2016})}\BibitemShut {NoStop}%
\bibitem [{\citenamefont {Tsvelick}\ and\ \citenamefont
  {Wiegmann}(1983)}]{Tsvelick_1983_R}%
  \BibitemOpen
  \bibfield  {author} {\bibinfo {author} {\bibfnamefont {A.M.}\ \bibnamefont
  {Tsvelick}}\ and\ \bibinfo {author} {\bibfnamefont {P.B.}\ \bibnamefont
  {Wiegmann}},\ }\bibfield  {title} {\enquote {\bibinfo {title} {Exact results
  in the theory of magnetic alloys},}\ }\href {\doibase
  10.1080/00018738300101581} {\bibfield  {journal} {\bibinfo  {journal}
  {Advances in Physics}\ }\textbf {\bibinfo {volume} {32}},\ \bibinfo {pages}
  {453} (\bibinfo {year} {1983})}\BibitemShut {NoStop}%
\bibitem [{\citenamefont {Okiji}\ and\ \citenamefont
  {Kawakami}(1984)}]{Okiji_1984}%
  \BibitemOpen
  \bibfield  {author} {\bibinfo {author} {\bibfnamefont {Ayao}\ \bibnamefont
  {Okiji}}\ and\ \bibinfo {author} {\bibfnamefont {Norio}\ \bibnamefont
  {Kawakami}},\ }\bibfield  {title} {\enquote {\bibinfo {title} {Thermodynamic
  properties of the {Anderson} model (invited)},}\ }\href {\doibase
  10.1063/1.333523} {\bibfield  {journal} {\bibinfo  {journal} {Journal of
  Applied Physics}\ }\textbf {\bibinfo {volume} {55}},\ \bibinfo {pages} {1931}
  (\bibinfo {year} {1984})}\BibitemShut {NoStop}%
\bibitem [{\citenamefont {Tans}\ \emph {et~al.}(1997)\citenamefont {Tans},
  \citenamefont {Devoret}, \citenamefont {Dai}, \citenamefont {Thess},
  \citenamefont {Smalley}, \citenamefont {Geerligs},\ and\ \citenamefont
  {Dekker}}]{Tans_1997}%
  \BibitemOpen
  \bibfield  {author} {\bibinfo {author} {\bibfnamefont {Sander~J.}\
  \bibnamefont {Tans}}, \bibinfo {author} {\bibfnamefont {Michel~H.}\
  \bibnamefont {Devoret}}, \bibinfo {author} {\bibfnamefont {Hongjie}\
  \bibnamefont {Dai}}, \bibinfo {author} {\bibfnamefont {Andreas}\ \bibnamefont
  {Thess}}, \bibinfo {author} {\bibfnamefont {Richard~E.}\ \bibnamefont
  {Smalley}}, \bibinfo {author} {\bibfnamefont {L.~J.}\ \bibnamefont
  {Geerligs}}, \ and\ \bibinfo {author} {\bibfnamefont {Cees}\ \bibnamefont
  {Dekker}},\ }\bibfield  {title} {\enquote {\bibinfo {title} {Individual
  single-wall carbon nanotubes as quantum wires},}\ }\href {\doibase
  10.1038/386474a0} {\bibfield  {journal} {\bibinfo  {journal} {Nature}\
  }\textbf {\bibinfo {volume} {386}},\ \bibinfo {pages} {474} (\bibinfo {year}
  {1997})}\BibitemShut {NoStop}%
\bibitem [{\citenamefont {Nygård}\ \emph {et~al.}(2000)\citenamefont
  {Nygård}, \citenamefont {Cobden},\ and\ \citenamefont
  {Lindelof}}]{Nygard_2000}%
  \BibitemOpen
  \bibfield  {author} {\bibinfo {author} {\bibfnamefont {Jesper}\ \bibnamefont
  {Nygård}}, \bibinfo {author} {\bibfnamefont {David~Henry}\ \bibnamefont
  {Cobden}}, \ and\ \bibinfo {author} {\bibfnamefont {Poul~Erik}\ \bibnamefont
  {Lindelof}},\ }\bibfield  {title} {\enquote {\bibinfo {title} {Kondo physics
  in carbon nanotubes},}\ }\href {\doibase 10.1038/35042545} {\bibfield
  {journal} {\bibinfo  {journal} {Nature}\ }\textbf {\bibinfo {volume} {408}},\
  \bibinfo {pages} {342} (\bibinfo {year} {2000})}\BibitemShut {NoStop}%
\bibitem [{\citenamefont {Liang}\ \emph {et~al.}(2001)\citenamefont {Liang},
  \citenamefont {Bockrath}, \citenamefont {Bozovic}, \citenamefont {Hafner},
  \citenamefont {Tinkham},\ and\ \citenamefont {Park}}]{Liang_2001}%
  \BibitemOpen
  \bibfield  {author} {\bibinfo {author} {\bibfnamefont {Wenjie}\ \bibnamefont
  {Liang}}, \bibinfo {author} {\bibfnamefont {Marc}\ \bibnamefont {Bockrath}},
  \bibinfo {author} {\bibfnamefont {Dolores}\ \bibnamefont {Bozovic}}, \bibinfo
  {author} {\bibfnamefont {Jason~H.}\ \bibnamefont {Hafner}}, \bibinfo {author}
  {\bibfnamefont {M.}~\bibnamefont {Tinkham}}, \ and\ \bibinfo {author}
  {\bibfnamefont {Hongkun}\ \bibnamefont {Park}},\ }\bibfield  {title}
  {\enquote {\bibinfo {title} {Fabry - {Perot} interference in a nanotube
  electron waveguide},}\ }\href {\doibase 10.1038/35079517} {\bibfield
  {journal} {\bibinfo  {journal} {Nature}\ }\textbf {\bibinfo {volume} {411}},\
  \bibinfo {pages} {665} (\bibinfo {year} {2001})}\BibitemShut {NoStop}%
\bibitem [{\citenamefont {Roch}\ \emph {et~al.}(2008)\citenamefont {Roch},
  \citenamefont {Florens}, \citenamefont {Bouchiat}, \citenamefont
  {Wernsdorfer},\ and\ \citenamefont {Balestro}}]{Roch_2007}%
  \BibitemOpen
  \bibfield  {author} {\bibinfo {author} {\bibfnamefont {Nicolas}\ \bibnamefont
  {Roch}}, \bibinfo {author} {\bibfnamefont {Serge}\ \bibnamefont {Florens}},
  \bibinfo {author} {\bibfnamefont {Vincent}\ \bibnamefont {Bouchiat}},
  \bibinfo {author} {\bibfnamefont {Wolfgang}\ \bibnamefont {Wernsdorfer}}, \
  and\ \bibinfo {author} {\bibfnamefont {Franck}\ \bibnamefont {Balestro}},\
  }\bibfield  {title} {\enquote {\bibinfo {title} {Quantum phase transition in
  a single-molecule quantum dot},}\ }\href {\doibase 10.1038/nature06930}
  {\bibfield  {journal} {\bibinfo  {journal} {Nature}\ }\textbf {\bibinfo
  {volume} {453}},\ \bibinfo {pages} {633} (\bibinfo {year}
  {2008})}\BibitemShut {NoStop}%
\bibitem [{\citenamefont {Ridley}\ \emph {et~al.}(2019)\citenamefont {Ridley},
  \citenamefont {Galperin}, \citenamefont {Gull},\ and\ \citenamefont
  {Cohen}}]{Ridley_1907}%
  \BibitemOpen
  \bibfield  {author} {\bibinfo {author} {\bibfnamefont {Michael}\ \bibnamefont
  {Ridley}}, \bibinfo {author} {\bibfnamefont {Michael}\ \bibnamefont
  {Galperin}}, \bibinfo {author} {\bibfnamefont {Emanuel}\ \bibnamefont
  {Gull}}, \ and\ \bibinfo {author} {\bibfnamefont {Guy}\ \bibnamefont
  {Cohen}},\ }\bibfield  {title} {\enquote {\bibinfo {title} {Numerically exact
  full counting statistics of the energy current in the {Kondo} regime},}\
  }\href {\doibase 10.1103/PhysRevB.100.165127} {\bibfield  {journal} {\bibinfo
   {journal} {Phys. Rev. B}\ }\textbf {\bibinfo {volume} {100}},\ \bibinfo
  {pages} {165127} (\bibinfo {year} {2019})},\ \Eprint
  {http://arxiv.org/abs/1907.09546} {arXiv:1907.09546} \BibitemShut {NoStop}%
\bibitem [{\citenamefont {Krivenko}\ \emph {et~al.}(2019)\citenamefont
  {Krivenko}, \citenamefont {Kleinhenz}, \citenamefont {Cohen},\ and\
  \citenamefont {Gull}}]{Krivenko_1904}%
  \BibitemOpen
  \bibfield  {author} {\bibinfo {author} {\bibfnamefont {Igor}\ \bibnamefont
  {Krivenko}}, \bibinfo {author} {\bibfnamefont {Joseph}\ \bibnamefont
  {Kleinhenz}}, \bibinfo {author} {\bibfnamefont {Guy}\ \bibnamefont {Cohen}},
  \ and\ \bibinfo {author} {\bibfnamefont {Emanuel}\ \bibnamefont {Gull}},\
  }\bibfield  {title} {\enquote {\bibinfo {title} {Dynamics of {Kondo} voltage
  splitting after a quantum quench},}\ }\href {\doibase
  10.1103/PhysRevB.100.201104} {\bibfield  {journal} {\bibinfo  {journal}
  {Phys. Rev. B}\ }\textbf {\bibinfo {volume} {100}},\ \bibinfo {pages}
  {201104} (\bibinfo {year} {2019})},\ \Eprint
  {http://arxiv.org/abs/1904.11527} {arXiv:1904.11527} \BibitemShut {NoStop}%
\bibitem [{\citenamefont {Beenakker}(1991)}]{Beenakker_1990}%
  \BibitemOpen
  \bibfield  {author} {\bibinfo {author} {\bibfnamefont {C.~W.~J.}\
  \bibnamefont {Beenakker}},\ }\bibfield  {title} {\enquote {\bibinfo {title}
  {Theory of {Coulomb}-blockade oscillations in the conductance of a quantum
  dot},}\ }\href {\doibase 10.1103/PhysRevB.44.1646} {\bibfield  {journal}
  {\bibinfo  {journal} {Phys. Rev. B}\ }\textbf {\bibinfo {volume} {44}},\
  \bibinfo {pages} {1646} (\bibinfo {year} {1991})}\BibitemShut {NoStop}%
\bibitem [{\citenamefont {Hofheinz}\ \emph {et~al.}(2007)\citenamefont
  {Hofheinz}, \citenamefont {Jehl}, \citenamefont {Sanquer}, \citenamefont
  {Molas}, \citenamefont {Vinet},\ and\ \citenamefont
  {Deleonibus}}]{Hofheinz_2007}%
  \BibitemOpen
  \bibfield  {author} {\bibinfo {author} {\bibfnamefont {M.}~\bibnamefont
  {Hofheinz}}, \bibinfo {author} {\bibfnamefont {X.}~\bibnamefont {Jehl}},
  \bibinfo {author} {\bibfnamefont {M.}~\bibnamefont {Sanquer}}, \bibinfo
  {author} {\bibfnamefont {G.}~\bibnamefont {Molas}}, \bibinfo {author}
  {\bibfnamefont {M.}~\bibnamefont {Vinet}}, \ and\ \bibinfo {author}
  {\bibfnamefont {S.}~\bibnamefont {Deleonibus}},\ }\bibfield  {title}
  {\enquote {\bibinfo {title} {Capacitance enhancement in {Coulomb} blockade
  tunnel barriers},}\ }\href {\doibase 10.1103/PhysRevB.75.235301} {\bibfield
  {journal} {\bibinfo  {journal} {Phys. Rev. B}\ }\textbf {\bibinfo {volume}
  {75}},\ \bibinfo {pages} {235301} (\bibinfo {year} {2007})}\BibitemShut
  {NoStop}%
\bibitem [{\citenamefont {Schollwöck}(2011)}]{Schollwoeck_1008}%
  \BibitemOpen
  \bibfield  {author} {\bibinfo {author} {\bibfnamefont {Ulrich}\ \bibnamefont
  {Schollwöck}},\ }\bibfield  {title} {\enquote {\bibinfo {title} {The
  density-matrix renormalization group in the age of matrix product states},}\
  }\href {\doibase 10.1016/j.aop.2010.09.012} {\bibfield  {journal} {\bibinfo
  {journal} {Annals of Physics}\ }\textbf {\bibinfo {volume} {326}},\ \bibinfo
  {pages} {96} (\bibinfo {year} {2011})},\ \Eprint
  {http://arxiv.org/abs/1008.3477} {arXiv:1008.3477} \BibitemShut {NoStop}%
\bibitem [{\citenamefont {Glasser}\ \emph {et~al.}()\citenamefont {Glasser},
  \citenamefont {Sweke}, \citenamefont {Pancotti}, \citenamefont {Eisert},\
  and\ \citenamefont {Cirac}}]{Glasser_1907}%
  \BibitemOpen
  \bibfield  {author} {\bibinfo {author} {\bibfnamefont {Ivan}\ \bibnamefont
  {Glasser}}, \bibinfo {author} {\bibfnamefont {Ryan}\ \bibnamefont {Sweke}},
  \bibinfo {author} {\bibfnamefont {Nicola}\ \bibnamefont {Pancotti}}, \bibinfo
  {author} {\bibfnamefont {Jens}\ \bibnamefont {Eisert}}, \ and\ \bibinfo
  {author} {\bibfnamefont {J.~Ignacio}\ \bibnamefont {Cirac}},\ }\bibfield
  {title} {\enquote {\bibinfo {title} {Expressive power of tensor-network
  factorizations for probabilistic modeling, with applications from hidden
  {Markov} models to quantum machine learning},}\ }\href@noop {} {\ }\Eprint
  {http://arxiv.org/abs/1907.03741} {arXiv:1907.03741} \BibitemShut {NoStop}%
\bibitem [{\citenamefont {Parcollet}\ \emph {et~al.}(2015)\citenamefont
  {Parcollet}, \citenamefont {Ferrero}, \citenamefont {Ayral}, \citenamefont
  {Hafermann}, \citenamefont {Krivenko}, \citenamefont {Messio},\ and\
  \citenamefont {Seth}}]{TRIQS2015}%
  \BibitemOpen
  \bibfield  {author} {\bibinfo {author} {\bibfnamefont {Olivier}\ \bibnamefont
  {Parcollet}}, \bibinfo {author} {\bibfnamefont {Michel}\ \bibnamefont
  {Ferrero}}, \bibinfo {author} {\bibfnamefont {Thomas}\ \bibnamefont {Ayral}},
  \bibinfo {author} {\bibfnamefont {Hartmut}\ \bibnamefont {Hafermann}},
  \bibinfo {author} {\bibfnamefont {Igor}\ \bibnamefont {Krivenko}}, \bibinfo
  {author} {\bibfnamefont {Laura}\ \bibnamefont {Messio}}, \ and\ \bibinfo
  {author} {\bibfnamefont {Priyanka}\ \bibnamefont {Seth}},\ }\bibfield
  {title} {\enquote {\bibinfo {title} {{TRIQS}: A toolbox for research on
  interacting quantum systems},}\ }\href {\doibase
  http://dx.doi.org/10.1016/j.cpc.2015.04.023} {\bibfield  {journal} {\bibinfo
  {journal} {Computer Physics Communications}\ }\textbf {\bibinfo {volume}
  {196}},\ \bibinfo {pages} {398 -- 415} (\bibinfo {year} {2015})}\BibitemShut
  {NoStop}%
\bibitem [{\citenamefont {Roth}(1954)}]{Roth_1954}%
  \BibitemOpen
  \bibfield  {author} {\bibinfo {author} {\bibfnamefont {K.~F.}\ \bibnamefont
  {Roth}},\ }\bibfield  {title} {\enquote {\bibinfo {title} {On irregularities
  of distribution},}\ }\href {\doibase 10.1112/S0025579300000541} {\bibfield
  {journal} {\bibinfo  {journal} {Mathematika}\ }\textbf {\bibinfo {volume}
  {1}},\ \bibinfo {pages} {73--79} (\bibinfo {year} {1954})}\BibitemShut
  {NoStop}%
\bibitem [{\citenamefont {Weyl}(1916)}]{Weyl_1916}%
  \BibitemOpen
  \bibfield  {author} {\bibinfo {author} {\bibfnamefont {Hermann}\ \bibnamefont
  {Weyl}},\ }\bibfield  {title} {\enquote {\bibinfo {title} {{\"U}ber die
  {Gleichverteilung} von {Zahlen} mod. {Eins}},}\ }\href {\doibase
  10.1007/BF01475864} {\bibfield  {journal} {\bibinfo  {journal} {Mathematische
  Annalen}\ }\textbf {\bibinfo {volume} {77}},\ \bibinfo {pages} {313--352}
  (\bibinfo {year} {1916})}\BibitemShut {NoStop}%
\bibitem [{\citenamefont {Korobov}(1957)}]{Korobov_1957}%
  \BibitemOpen
  \bibfield  {author} {\bibinfo {author} {\bibfnamefont {N.~M.}\ \bibnamefont
  {Korobov}},\ }\bibfield  {title} {\enquote {\bibinfo {title} {Approximate
  calculation of repeated integrals by number-theoretical methods},}\ }\href
  {http://mi.mathnet.ru/eng/dan/v115/i6/p1062} {\bibfield  {journal} {\bibinfo
  {journal} {Doklady Akademii Nauk SSSR}\ }\textbf {\bibinfo {volume} {115}},\
  \bibinfo {pages} {1062--1065} (\bibinfo {year} {1957})}\BibitemShut {NoStop}%
\bibitem [{\citenamefont {Hammersley}(1960)}]{Hammersley_1960}%
  \BibitemOpen
  \bibfield  {author} {\bibinfo {author} {\bibfnamefont {J.~M.}\ \bibnamefont
  {Hammersley}},\ }\bibfield  {title} {\enquote {\bibinfo {title} {Monte
  {Carlo} methods for solving multivariable problems},}\ }\href {\doibase
  10.1111/j.1749-6632.1960.tb42846.x} {\bibfield  {journal} {\bibinfo
  {journal} {Annals of the New York Academy of Sciences}\ }\textbf {\bibinfo
  {volume} {86}},\ \bibinfo {pages} {844--874} (\bibinfo {year}
  {1960})}\BibitemShut {NoStop}%
\bibitem [{\citenamefont {van~der Corput}(1935)}]{Corput_1935}%
  \BibitemOpen
  \bibfield  {author} {\bibinfo {author} {\bibfnamefont {J.~C.}\ \bibnamefont
  {van~der Corput}},\ }\bibfield  {title} {\enquote {\bibinfo {title}
  {Verteilungsfunktionen},}\ }\href@noop {} {\bibfield  {journal} {\bibinfo
  {journal} {Nederl. Akad. Wetensch. Proc.}\ }\textbf {\bibinfo {volume}
  {38}},\ \bibinfo {pages} {813--821, 1058--1066} (\bibinfo {year}
  {1935})}\BibitemShut {NoStop}%
\bibitem [{\citenamefont {Halton}(1960)}]{Halton_1960}%
  \BibitemOpen
  \bibfield  {author} {\bibinfo {author} {\bibfnamefont {J.~H.}\ \bibnamefont
  {Halton}},\ }\bibfield  {title} {\enquote {\bibinfo {title} {On the
  efficiency of certain quasi-random sequences of points in evaluating
  multi-dimensional integrals},}\ }\href {\doibase 10.1007/BF01386213}
  {\bibfield  {journal} {\bibinfo  {journal} {Numerische Mathematik}\ }\textbf
  {\bibinfo {volume} {2}},\ \bibinfo {pages} {84--90} (\bibinfo {year}
  {1960})}\BibitemShut {NoStop}%
\bibitem [{\citenamefont {Haselgrove}(1961)}]{Haselgrove_1961}%
  \BibitemOpen
  \bibfield  {author} {\bibinfo {author} {\bibfnamefont {C.~B.}\ \bibnamefont
  {Haselgrove}},\ }\bibfield  {title} {\enquote {\bibinfo {title} {A method for
  numerical integration},}\ }\href {\doibase 10.1090/S0025-5718-1961-0146960-1}
  {\bibfield  {journal} {\bibinfo  {journal} {Mathematics of Computation}\
  }\textbf {\bibinfo {volume} {15}},\ \bibinfo {pages} {323--337} (\bibinfo
  {year} {1961})}\BibitemShut {NoStop}%
\bibitem [{\citenamefont {Richtmyer}(1952)}]{Richtmyer_1952}%
  \BibitemOpen
  \bibfield  {author} {\bibinfo {author} {\bibfnamefont {R~D}\ \bibnamefont
  {Richtmyer}},\ }\href {\doibase 10.2172/4405295} {\emph {\bibinfo {title}
  {The evaluation of definite integrals, and a quasi-{Monte}-{Carlo} method
  based on the properties of algebraic numbers}}},\ \bibinfo {type} {Tech.
  Rep.}\ \bibinfo {number} {LA-1342}\ (\bibinfo  {institution} {Los Alamos
  Scientific Lab.},\ \bibinfo {address} {Los Alamos, NM},\ \bibinfo {year}
  {1952})\BibitemShut {NoStop}%
\bibitem [{\citenamefont {Niederreiter}(1978)}]{Niederreiter_1978}%
  \BibitemOpen
  \bibfield  {author} {\bibinfo {author} {\bibfnamefont {Harald}\ \bibnamefont
  {Niederreiter}},\ }\bibfield  {title} {\enquote {\bibinfo {title}
  {Quasi-{Monte} {Carlo} methods and pseudo-random numbers},}\ }\href {\doibase
  10.1090/S0002-9904-1978-14532-7} {\bibfield  {journal} {\bibinfo  {journal}
  {Bulletin of the American Mathematical Society}\ }\textbf {\bibinfo {volume}
  {84}},\ \bibinfo {pages} {957--1041} (\bibinfo {year} {1978})}\BibitemShut
  {NoStop}%
\bibitem [{\citenamefont {Conroy}(1967)}]{Conroy_1967}%
  \BibitemOpen
  \bibfield  {author} {\bibinfo {author} {\bibfnamefont {Harold}\ \bibnamefont
  {Conroy}},\ }\bibfield  {title} {\enquote {\bibinfo {title} {Molecular
  {Schrödinger} equation. {VIII}. {A} new method for the evaluation of
  multidimensional integrals},}\ }\href {\doibase 10.1063/1.1701795} {\bibfield
   {journal} {\bibinfo  {journal} {The Journal of Chemical Physics}\ }\textbf
  {\bibinfo {volume} {47}},\ \bibinfo {pages} {5307--5318} (\bibinfo {year}
  {1967})}\BibitemShut {NoStop}%
\bibitem [{\citenamefont {Berblinger}\ and\ \citenamefont
  {Schlier}(1991)}]{Berblinger_1991}%
  \BibitemOpen
  \bibfield  {author} {\bibinfo {author} {\bibfnamefont {Michael}\ \bibnamefont
  {Berblinger}}\ and\ \bibinfo {author} {\bibfnamefont {Christoph}\
  \bibnamefont {Schlier}},\ }\bibfield  {title} {\enquote {\bibinfo {title}
  {Monte {Carlo} integration with quasi-random numbers: some experience},}\
  }\href {\doibase 10.1016/0010-4655(91)90064-R} {\bibfield  {journal}
  {\bibinfo  {journal} {Computer Physics Communications}\ }\textbf {\bibinfo
  {volume} {66}},\ \bibinfo {pages} {157--166} (\bibinfo {year}
  {1991})}\BibitemShut {NoStop}%
\bibitem [{\citenamefont {Paskov}\ and\ \citenamefont
  {Traub}(1995)}]{Paskov_1995}%
  \BibitemOpen
  \bibfield  {author} {\bibinfo {author} {\bibfnamefont {Spassimir~H.}\
  \bibnamefont {Paskov}}\ and\ \bibinfo {author} {\bibfnamefont {Joseph~F.}\
  \bibnamefont {Traub}},\ }\bibfield  {title} {\enquote {\bibinfo {title}
  {Faster {Valuation} of {Financial} {Derivatives}},}\ }\href {\doibase
  10.3905/jpm.1995.409541} {\bibfield  {journal} {\bibinfo  {journal} {J.
  Portf. Manag.}\ }\textbf {\bibinfo {volume} {22}},\ \bibinfo {pages}
  {113--123} (\bibinfo {year} {1995})}\BibitemShut {NoStop}%
\bibitem [{\citenamefont {Wang}\ and\ \citenamefont {Sloan}(2005)}]{Wang_2005}%
  \BibitemOpen
  \bibfield  {author} {\bibinfo {author} {\bibfnamefont {Xiaoqun}\ \bibnamefont
  {Wang}}\ and\ \bibinfo {author} {\bibfnamefont {Ian~H.}\ \bibnamefont
  {Sloan}},\ }\bibfield  {title} {\enquote {\bibinfo {title} {Why are
  high-dimensional finance problems often of low effective dimension?}}\ }\href
  {\doibase 10.1137/S1064827503429429} {\bibfield  {journal} {\bibinfo
  {journal} {SIAM Journal on Scientific Computing}\ }\textbf {\bibinfo {volume}
  {27}},\ \bibinfo {pages} {159--183} (\bibinfo {year} {2005})}\BibitemShut
  {NoStop}%
\bibitem [{\citenamefont {Nuyens}\ and\ \citenamefont
  {Cools}(2006)}]{Nuyens_2006}%
  \BibitemOpen
  \bibfield  {author} {\bibinfo {author} {\bibfnamefont {Dirk}\ \bibnamefont
  {Nuyens}}\ and\ \bibinfo {author} {\bibfnamefont {Ronald}\ \bibnamefont
  {Cools}},\ }\bibfield  {title} {\enquote {\bibinfo {title} {Fast algorithms
  for component-by-component construction of rank-1 lattice rules in
  shift-invariant reproducing kernel {Hilbert} spaces},}\ }\href {\doibase
  10.1090/S0025-5718-06-01785-6} {\bibfield  {journal} {\bibinfo  {journal}
  {Mathematics of Computation}\ }\textbf {\bibinfo {volume} {75}},\ \bibinfo
  {pages} {903--920} (\bibinfo {year} {2006})}\BibitemShut {NoStop}%
\bibitem [{\citenamefont {Lepage}(1978)}]{Lepage_1978}%
  \BibitemOpen
  \bibfield  {author} {\bibinfo {author} {\bibfnamefont {G.~Peter}\
  \bibnamefont {Lepage}},\ }\bibfield  {title} {\enquote {\bibinfo {title} {A
  new algorithm for adaptive multidimensional integration},}\ }\href {\doibase
  10.1016/0021-9991(78)90004-9} {\bibfield  {journal} {\bibinfo  {journal}
  {Journal of Computational Physics}\ }\textbf {\bibinfo {volume} {27}},\
  \bibinfo {pages} {192} (\bibinfo {year} {1978})}\BibitemShut {NoStop}%
\bibitem [{\citenamefont {Lepage}(1980)}]{Lepage_1980}%
  \BibitemOpen
  \bibfield  {author} {\bibinfo {author} {\bibfnamefont {G.~Peter}\
  \bibnamefont {Lepage}},\ }\href@noop {} {\emph {\bibinfo {title} {{VEGAS - an
  adaptive multi-dimensional integration program}}}},\ \bibinfo {type} {Tech.
  Rep.}\ \bibinfo {number} {CLNS-447}\ (\bibinfo  {institution} {Cornell Univ.
  Lab. Nucl. Stud.},\ \bibinfo {address} {Ithaca, NY},\ \bibinfo {year}
  {1980})\BibitemShut {NoStop}%
\bibitem [{\citenamefont {Rubtsov}\ and\ \citenamefont
  {Lichtenstein}(2004)}]{Rubtsov2004}%
  \BibitemOpen
  \bibfield  {author} {\bibinfo {author} {\bibfnamefont {A.~N.}\ \bibnamefont
  {Rubtsov}}\ and\ \bibinfo {author} {\bibfnamefont {A.~I.}\ \bibnamefont
  {Lichtenstein}},\ }\bibfield  {title} {\enquote {\bibinfo {title}
  {Continuous-time quantum {Monte} {Carlo} method for fermions: Beyond
  auxiliary field framework},}\ }\href {\doibase 10.1134/1.1800216} {\bibfield
  {journal} {\bibinfo  {journal} {Journal of Experimental and Theoretical
  Physics Letters}\ }\textbf {\bibinfo {volume} {80}},\ \bibinfo {pages}
  {61--65} (\bibinfo {year} {2004})}\BibitemShut {NoStop}%
\bibitem [{\citenamefont {Meir}\ and\ \citenamefont
  {Wingreen}(1992)}]{Meir_1992}%
  \BibitemOpen
  \bibfield  {author} {\bibinfo {author} {\bibfnamefont {Yigal}\ \bibnamefont
  {Meir}}\ and\ \bibinfo {author} {\bibfnamefont {Ned~S.}\ \bibnamefont
  {Wingreen}},\ }\bibfield  {title} {\enquote {\bibinfo {title} {Landauer
  formula for the current through an interacting electron region},}\ }\href
  {\doibase 10.1103/PhysRevLett.68.2512} {\bibfield  {journal} {\bibinfo
  {journal} {Phys. Rev. Lett.}\ }\textbf {\bibinfo {volume} {68}},\ \bibinfo
  {pages} {2512--2515} (\bibinfo {year} {1992})}\BibitemShut {NoStop}%
\bibitem [{\citenamefont {Wiegmann}\ and\ \citenamefont
  {Tsvelick}(1983)}]{Wiegmann_1983_a}%
  \BibitemOpen
  \bibfield  {author} {\bibinfo {author} {\bibfnamefont {P~B}\ \bibnamefont
  {Wiegmann}}\ and\ \bibinfo {author} {\bibfnamefont {A~M}\ \bibnamefont
  {Tsvelick}},\ }\bibfield  {title} {\enquote {\bibinfo {title} {Exact solution
  of the {Anderson} model: I},}\ }\href {\doibase 10.1088/0022-3719/16/12/017}
  {\bibfield  {journal} {\bibinfo  {journal} {J. Phys. C: Solid State Phys.}\
  }\textbf {\bibinfo {volume} {16}},\ \bibinfo {pages} {2281} (\bibinfo {year}
  {1983})}\BibitemShut {NoStop}%
\bibitem [{\citenamefont {Horvati{\'c}}\ and\ \citenamefont
  {Zlati{\'c}}(1985)}]{Horvatic_1985}%
  \BibitemOpen
  \bibfield  {author} {\bibinfo {author} {\bibfnamefont {B.}~\bibnamefont
  {Horvati{\'c}}}\ and\ \bibinfo {author} {\bibfnamefont {V.}~\bibnamefont
  {Zlati{\'c}}},\ }\bibfield  {title} {\enquote {\bibinfo {title} {Equivalence
  of the perturbative and {Bethe}-{Ansatz} solution of the symmetric {Anderson}
  {Hamiltonian}},}\ }\href {\doibase 10.1051/jphys:019850046090145900}
  {\bibfield  {journal} {\bibinfo  {journal} {J. Phys. France}\ }\textbf
  {\bibinfo {volume} {46}},\ \bibinfo {pages} {1459--1467} (\bibinfo {year}
  {1985})}\BibitemShut {NoStop}%
\end{thebibliography}%

\end{document}